\documentclass[12pt]{article}
\pdfoutput=1

\usepackage{amsmath}
\usepackage{amssymb}
\usepackage{epsfig}
\usepackage{graphicx}
\usepackage{cite}
\usepackage{multirow}
\usepackage{longtable}
\usepackage{lscape}
\usepackage{bbm}
\usepackage{mathrsfs}

\allowdisplaybreaks[4] 

\newcommand{\capdef}{}
\newcommand{\mycaption}[2][\capdef]{\renewcommand{\capdef}{#2}%
        \caption[#1]{{\footnotesize #2}}}
\makeatletter
\renewcommand{\fnum@table}{\textbf{\tablename~\thetable}}
\renewcommand{\fnum@figure}{\textbf{\figurename~\thefigure}}
\makeatother

\newcounter{myenumi}

\renewcommand{\themyenumi}{\roman{myenumi}}
{\end{list}}

\newlength{\myem}
\newcounter{mysubequation}[equation]

\makeatletter
\renewcommand{\section}{\@startsection{section}{1}{0em}{-\baselineskip}%
{\baselineskip}{\normalfont\large\bfseries}}
\renewcommand{\subsection}%
{\@startsection{subsection}{2}{0em}{-0.7\baselineskip}%
{0.7\baselineskip}{\normalfont\bfseries}}
\makeatother

\newcommand{\bi}{\begin{itemize}}
\newcommand{\ei}{\end{itemize}}

\newcommand{\be}{\begin{equation}}
\newcommand{\ee}{\end{equation}}
\newcommand{\bea}{\begin{eqnarray}}
\newcommand{\eea}{\end{eqnarray}}

\newcommand{\ldm}{\Delta m_{31}^2}
\newcommand{\sdm}{\Delta m_{21}^2}
\newcommand{\deltacp}{\delta_{\mathrm{CP}}}
\newcommand{\stheta}{\sin^2 2 \theta_{13}}

\newcommand{\ie}{{\it i.e.}}

\newcommand{\eg}{{\it e.g.}}

\newcommand{\cf}{{\it cf.}}

\newcommand{\fig}{Fig.}

\newcommand{\Ref}{Ref.}
\newcommand{\Refs}{Refs.}
\newcommand{\Sec}{Sec.}

\newcommand{\figu}[1]{\fig~\ref{fig:#1}}

\begin{document}

\begin{titlepage}

\renewcommand{\thefootnote}{\alph{footnote}}

\vspace*{-3.cm}
\begin{flushright}
EURONU-WP6-09-12 \\
IDS-NF-010
\end{flushright}


\renewcommand{\thefootnote}{\fnsymbol{footnote}}
\setcounter{footnote}{-1}

{\begin{center}
{\large\bf
Neutrino factory in stages: Low energy, high energy, off-axis
} \end{center}}
\renewcommand{\thefootnote}{\alph{footnote}}

\vspace*{.8cm}
\vspace*{.3cm}
{\begin{center} {\large{\sc
 		Jian~Tang\footnote[1]{\makebox[1.cm]{Email:}
                jtang@physik.uni-wuerzburg.de},
                Walter~Winter\footnote[2]{\makebox[1.cm]{Email:}
                winter@physik.uni-wuerzburg.de}
                }}
\end{center}}
\vspace*{0cm}
{\it
\begin{center}

\footnotemark[1]${}^,$\footnotemark[2]
       Institut f{\"u}r Theoretische Physik und Astrophysik, Universit{\"a}t W{\"u}rzburg, \\
       D-97074 W{\"u}rzburg, Germany

\end{center}}

\vspace*{1.5cm}

{\Large \bf
\begin{center} Abstract \end{center}  }

We discuss neutrino oscillation physics with a neutrino factory in stages,
including the possibility of upgrading the muon energy 
within the same program. We point out that a detector
designed for the low energy neutrino factory may be used off-axis in a 
high energy neutrino factory beam. We include 
the re-optimization of the experiment depending on the value of $\theta_{13}$ 
found. As upgrade options, we consider muon energy, additional baselines, a
detector mass upgrade, an off-axis detector, and the platinum (muon to electron neutrino)
channels. In addition, we test the impact of Daya Bay data on the optimization.
We find that for large $\theta_{13}$ ($\theta_{13}$ discovered by the next generation
of experiments), a low energy neutrino factory might be
the most plausible minimal version to test the unknown parameters. 
However, if a higher muon energy is needed for
new physics searches, a high energy version including an off-axis detector
may be an interesting alternative. For small $\theta_{13}$ ($\theta_{13}$ not
discovered by the next generation), a plausible
program could start with a low energy neutrino factory, followed by energy
upgrade, and then baseline or detector mass upgrade, depending on the
outcome of the earlier phases. 

\vspace*{.5cm}

\end{titlepage}

\newpage

\renewcommand{\thefootnote}{\arabic{footnote}}
\setcounter{footnote}{0}

\section{Introduction}

Three-flavor neutrino oscillations have been accepted as the most successful 
interpretation of neutrino flavor changes, 
see, \eg, \Ref~\cite{GonzalezGarcia:2007ib}.
In particular, the solar and atmospheric oscillation parameters have been
measured with very high precisions, and the reactor mixing angle $\theta_{13}$
has been strongly constrained. Future long-baseline and reactor neutrino 
experiments will test this small angle
further, and be sensitive to leptonic CP violation and the neutrino mass
hierarchy (see \Ref~\cite{Bandyopadhyay:2007kx} and references therein). The ultimate
instrument for these purposes might be a neutrino factory~\cite{Geer:1998iz,DeRujula:1998hd,Barger:1999fs,Cervera:2000kp}.
Using different baselines and oscillation channels, it can basically disentangle
all of the remaining oscillation parameters~\cite{Donini:2002rm,Autiero:2003fu,Huber:2003ak,Huber:2006wb} 
in spite of the presence of intrinsic correlations and degeneracies~\cite{Fogli:1996pv,Cervera:2000kp,Minakata:2001qm,Huber:2002mx}
for extremely small values of $\theta_{13}$.
Furthermore, a neutrino factory and other future neutrino oscillation experiments will
be sensitive to new physics searches, see \Ref~\cite{Bandyopadhyay:2007kx} for a summary.

The design of a neutrino factory has been put forward and discussed in international studies, such as in \Refs~\cite{Albright:2000xi, Apollonio:2002en,Albright:2004iw,Bandyopadhyay:2007kx}. Especially the most recent study, the International Neutrino Factory and Superbeam Scoping Study~\cite{Bandyopadhyay:2007kx,Abe:2007bi,Berg:2008xx}, has laid the foundations for the currently ongoing Design Study for the Neutrino Factory (IDS-NF)~\cite{ids}. This
initiative from about 2007 to 2012 is aiming to present a design report, schedule, cost estimate, and risk assessment for a neutrino factory. It defines a baseline setup of a high energy neutrino factory (HENF) with $E_\mu=25 \, \mathrm{GeV}$ and two baselines
$L_1 \simeq 4 \, 000 \, \mathrm{km}$ and $L_2 \simeq 7 \, 500 \, \mathrm{km}$ (the ``magic'' baseline) operated by two racetrack-shaped storage rings, where the muon energy is 25~GeV (for optimization questions, see \Refs~\cite{Barger:1999fs,Cervera:2000kp,Burguet-Castell:2001ez,Freund:2001ui,Donini:2005db,Huber:2006wb,Gandhi:2006gu,Kopp:2008ds}).  A key component is the magnetized iron detector (MIND) as far detector, where the magnetization is necessary to distinguish the ``right-sign'' (\eg, from $\nu_\mu \rightarrow \nu_\mu$) from the ``wrong-sign'' (\eg, from $\bar\nu_e \rightarrow \bar\nu_\mu$) muons. The identification of the muon charge of the wrong-sign muons allows, for example, for CP violation measurements in the muon neutrino appearance channels~\cite{Cervera:2000kp,Burguet-Castell:2001ez}. Possible near detector configurations for cross section and flux measurements have been discussed in \Refs~\cite{Abe:2007bi,Tang:2009na}.

As a more recent development, a low energy neutrino factory (LENF) with $E_\mu \simeq 4 \, \mathrm{GeV}$ to $5 \, \mathrm{GeV}$ has been proposed as an alternative to the HENF~\cite{Geer:2007kn,Bross:2007ts,Huber:2007uj,Bross:2009gk}. The main purpose of this alternative has been the reduction of accelerator cost in the case of large $\theta_{13}$. While the high energy neutrino factory relies on the MIND, the low energy neutrino factory as been proposed with a magnetized Totally Active Scintillator Detector (TASD), which allows for a lower threshold, better energy resolution, and (possibly) electron charge identification, which is required for the so-called ``platinum'' ($\nu_\mu \rightarrow \nu_e$ and $\bar\nu_\mu \rightarrow \bar\nu_e$) channels. These channels are the T-inverted channels of the muon neutrino appearance channels. Because of the same matter effects in the $\nu_\mu \rightarrow \nu_e$ and $\nu_e \rightarrow \nu_\mu$ (or $\bar\nu_\mu \rightarrow \bar\nu_e$ and $\bar\nu_e \rightarrow \bar\nu_\mu$)  channels, CP violation can, in principle, be extracted without convolution with the matter effects.\footnote{In the CP-conjugate channel, the matter effects are different, which means that the fundamental CP violation has to be disentangled from the Earth matter effects, which violate CP extrinsically (Earth matter does not contain any antimatter); see, \eg, \Ref~\cite{Akhmedov:2004ny}.} For the LENF, the useful number of muon decays may be increased by about 40\% by an optimization of the neutrino factory frontend for low energies~\cite{Bross:2009gk}.
Since the spectrum of a LENF can also be produced by a HENF in an off-axis detector (OAD), we consider this option as well. Such a detector would have a significant increase in low energy events.

As far as possible physics outcomes are concerned before the decision for a neutrino factory, we restrict ourselves to standard oscillation physics. The small mixing angle $\theta_{13}$ may be discovered by upcoming reactor (such as Double Chooz and Daya Bay) and superbeam (such as T2K and NO$\nu$A) experiments until about 2012 to 2015 if $\stheta \gtrsim 0.01$~\cite{Huber:2004ug,Huber:2009cw}, whereas for $\stheta \lesssim 0.01$, there will only be a new exclusion limit. At around this time, 
 a reference design report will be available by the IDS-NF~\cite{ids}, possibly with two different setups (a LENF and a HENF). In addition, LHC may have sufficient luminosity to indicate the new physics scale, which may point towards a muon collider, in favor of a neutrino factory. The decision for any future facility has to be based on this knowledge  (plus additional potential new physics discovered in oscillation and non-oscillation searches). We therefore refer to the $\stheta \simeq 0.01$ as the splitting point between the ``small $\theta_{13}$'' and ``large $\theta_{13}$'' cases, which marks the end of the reactor and (first generation) superbeam experiment dominated era. We use the terms ``$\theta_{13}$ not found by the next generation'' and ``$\theta_{13}$ found by the next generation'' equivalently. For the small $\theta_{13}$ case, the region $0.01 \lesssim \stheta \lesssim 0.1$ therefore is referred to as ``next generation excluded''. In our work, the large $\theta_{13}$ case is treated in \Sec~\ref{sec:largetheta13}, the small $\theta_{13}$ case in \Sec~\ref{sec:smalltheta13}. 

For several reasons, such as external boundary conditions, a neutrino factory complex may not be built at once, but instead be regarded as step-by-step program, perhaps, towards a muon collider.  In this study, we discuss both the LENF and HENF, where we are particularly interested in upgrade scenarios. This means that we demonstrate how building a neutrino factory in stages makes sense physics-wise, and we illustrate how the knowledge from earlier data affects the optimization. 
For the large $\theta_{13}$ case (\Sec~\ref{sec:largetheta13}), we discuss the minimal requirements for a neutrino factory setup to measure the yet unknown parameters for both LENF and HENF. In addition, we re-consider the baseline optimization for different matter density uncertainties,  the presence of the platinum channel or an off-axis detector, or the inclusion of Daya Bay data, which will be dominating the sensitivity to $\theta_{13}$ at that time~\cite{Huber:2009cw}. Furthermore, we show how much adding a second baseline or the combination between LENF and HENF will buy.  
For the small $\theta_{13}$ case (\Sec~\ref{sec:smalltheta13}), statistics in an off-axis detector will be very small, and platinum (because of charge identification backgrounds) or Daya Bay data (because of statistics) will not contribute significantly. In this case, we show a conceptually plausible upgrade scenario starting with a LENF, followed by an energy upgrade, and then by another detector upgrade or second baseline. We illustrate how the preceding phases affect the optimization, and we discuss possible synergies between LENF and HENF.
In the following \Sec~\ref{sec:exps}, we first of all motivate our simplifying assumptions, and we describe the experiments and possible upgrades.

\section{Assumptions and description of experiments and upgrades}
\label{sec:exps}

Here we describe our simplifying assumptions, experiments used, possible upgrades we will discuss, and the simulation techniques.

\subsection{Simplifying assumptions}

For the sake of simplicity, we make the following assumptions for this study:
\bi
\item
 We only discuss two different muon energies $E_\mu=4.12$~\cite{Bross:2007ts} (LENF) and $E_\mu=25 \, \mathrm{GeV}$~\cite{ids} (HENF).
The baseline-muon energy optimization has been performed in \Refs~\cite{Huber:2006wb,Huber:2007uj}. For all performance indicators, the conclusion has been that the desired performance can be reached with a certain threshold muon energy, but higher muon energies do not harm (within reasonable ranges). The muon energy choice of the HENF was driven by this observation, and $E_\mu=25 \, \mathrm{GeV}$ was as reasonable minimum for the MIND detector. In addition, note that the sensitivity to non-standard effects saturates at about this energy, because the matter resonance in the Earth's mantle can be covered at the peak~\cite{Kopp:2008ds}.
The LENF choice was motivated by the large $\theta_{13}$ performance given the TASD detector, with the boundary condition of low acceleration effort. More recently, somewhat larger $E_\mu \simeq 4.5$~\cite{Bross:2009gk} are considered, which do not affect this discussion qualitatively.
\item
We use the TASD as off-axis detector in the HENF with $E_\mu=25 \, \mathrm{GeV}$ with an off-axis angle of 0.55$^\circ$.
The considered spectrum of this off-axis neutrino factory detector (OAD) is practically identical to that of the LENF for this off-axis angle, which is a feature of the neutrino factory flux (see, \eg, \Ref~\cite{Tang:2009na}). Since the magnetized TASD is the detector used for the LENF and (so far) only discussed for that beam spectrum, we use the same beam spectrum. For $E_\mu=25 \, \mathrm{GeV}$, the corresponding off-axis angle is about 0.55$^\circ$, or about 38~km for $L=4 \, 000 \, \mathrm{km}$. In summary, we only use the MIND detector for HENF (on-axis), and the TASD for LENF (on-axis) and HENF (off-axis).
\item
 We normalize the luminosity of the neutrino beam produced by one decay straight to that of the IDS-NF baseline setup. The IDS-NF setup uses $2.5 \cdot 10^{20}$ useful muon decays per year, polarity, and decay straight, which corresponds to scale factor (SF) one. Since this setup uses two storage rings, the muons have to be shared among the rings. Therefore, if only one storage ring is used, such as for a one-baseline LENF or HENF, or a HENF+OAD, all muons can be injected in the same ring, and we have SF=2. If two baselines are operated simultaneously, on the other hand, we have SF=1. Note that the currently discussed LENF uses SF=2.8~\cite{Bross:2009gk}, where the additional 40\% increase comes from an optimization of the frontend.\footnote{These numbers are based on $10^7 \, \mathrm{s}$ operation time of the accelerator at the nominal luminosity per year. For  $2 \cdot 10^7 \, \mathrm{s}$ operation, one has SF=5.6 for the LENF, as the more aggressive setup in \Ref~\cite{Bross:2009gk}. We use $10^7 \, \mathrm{s}$ for all experiments in this study.}  Therefore, we sometimes show this as nominal scale factor. However, our approach has the advantage that LENF and HENF-OAD have the same performance. In addition, we require that the same frontend be used for LENF and HENF, which seems to make similar luminosities plausible. 
\item
 We assume that the platinum channel is only relevant for the TASD, \ie, LENF or HENF-OAD. The reason is that the electrons create showers, which leads to an upper threshold for platinum charge identification (and possibly unpredictable behavior).
\item
 We focus on standard three flavor oscillations with the three standard performance indicators: $\theta_{13}$ discovery, mass hierarchy (MH) discovery, and CP violation (CPV) discovery.
\end{itemize}

\subsection{Low energy neutrino factory and possible upgrades}

Our LENF is based upon \Ref~\cite{Bross:2007ts}.
It uses $E_\mu=4.12 \, \mathrm{GeV}$
and a magnetized Totally Active Scintillator Detector (TASD) 
with 20~kt fiducial mass times efficiency at (typically) 
$L=1300 \, \mathrm{km}$. We will re-consider the optimization
of the baseline including different types of upgrades.
The detector threshold is assumed to be $500 \, \mathrm{MeV}$
and the detection efficiency 73\%. Compared to \Ref~\cite{Bross:2007ts},
we use an energy resolution $\Delta E  = 0.1 \, \sqrt{E/\mathrm{GeV}} \, \mathrm{GeV}$, which is typical for TASD detectors such as NO$\nu$A. 
Both muon neutrino (and antineutrino) disappearance and appearance channels are included. The background level is conservatively estimated to be $10^{-3}$. Note that we include two types of backgrounds for the appearance channels, one which scales with the disappearance rates (such as from charge mis-identification), and one which scales with the un-oscillated spectrum (such as from neutral current events), both at the level of $10^{-3}$. As in  \Ref~\cite{Bross:2007ts}, we take the systematical errors to be 2\% for all signal and background errors. 

As potential ``upgrades'', these may the most relevant options:
\begin{description}
\item[Platinum channel]  Our description of the platinum $\nu_\mu \rightarrow \nu_e$ and $\bar\nu_\mu \rightarrow \bar\nu_e$ channels is based upon \Refs~\cite{NUMI714,Rubbia:2001pk,Huber:2006wb}. The efficiency is 40\%, the energy resolution is $\Delta E  = 0.15 \, E$, and the background level is about $0.01$. Note that the platinum channel is very difficult to use for higher energies, since the electrons induce showers for which the charge is difficult to measure. Therefore, we only use it in the TASD for the LENF (and HENF-OAD). Because of the relatively high background level, we only consider the platinum channel for large $\theta_{13}$ (\cf, \Ref~\cite{Huber:2006wb}).
\item[Energy upgrade] We consider an energy upgrade to the HENF (see below). Note that since the higher muon energies require a different detector technology, we use the MIND detector in that case.
\item[External $\boldsymbol{\theta_{13}}$ measurement] For large $\theta_{13}$, we consider the impact of an external measurement of $\theta_{13}$. Since we expect that Daya Bay dominates the $\theta_{13}$ sensitivity at the time given, we only use the data from this experiment simulated as in \Ref~\cite{Huber:2009cw}.
\end{description}

During the completion of this work, \Ref~\cite{Bross:2009gk} has appeared, which uses $E_\mu=4.5 \, \mathrm{GeV}$ and energy-dependent detection efficiencies. The background level for muon neutrino appearance and platinum channels in \Ref~\cite{Bross:2009gk} is effectively a factor of two smaller than in this study. The energy resolution is similar (10\%, linear in $E$). We have checked that the differences in efficiencies and energy resolution have practically no impact on the sensitivities. However, we conservatively assume higher background levels and a lower luminosity, which do have some impact. Therefore, our LENF has to be understood as conservative ``minimal'' choice LENF. As we will demonstrate, this choice is sufficient for the measurements for large $\stheta$.

\subsection{High energy neutrino factory and possible upgrades}

Our HENF is based upon the current IDS-NF baseline setup. In the standard configuration, it uses two 50~kt magnetized iron detectors (MIND) at baselines of about $4000 \, \mathrm{km}$ and $7500 \, \mathrm{km}$. We start with one baseline first ($4000 \, \mathrm{km}$ for small $\theta_{13}$), and then consider potential upgrades. Both $\nu_\mu$ disappearance and appearance channels are included. The energy resolution of the detectors is $\Delta E = 0.55 \sqrt{E / \mathrm{GeV}} \, \mathrm{GeV}$. The $\nu_\mu$ disappearance channels have a 90\% detection efficiency with a threshold of $1 \, \mathrm{GeV}$. The detector efficiencies (up to 74\%) and background fractions (between about $10^{-5}$ and $10^{-4}$) of the appearance channels are a function of the neutrino energy; for details, see \Ref~\cite{ids}. The systematical errors for signal and background normalizations are taken to be 2.5\%, uncorrelated among all channels. The number of useful muon decays per year, polarity, and storage ring is $2.5 \, 10^{20}$ (for two baselines operated simultaneously), which corresponds to SF=1. The simulation of the neutrino factory is based on \Refs~\cite{ids,Huber:2002mx,Huber:2006wb,Ables:1995wq}.

As potential ``upgrades'', we consider the following options:
\begin{description}
\item[Off-axis detector]
A potential OAD is considered at the off-axis angle 0.55$^\circ$, where the beam spectrum corresponds to the LENF spectrum. Therefore, the TASD from the LENF is used at the off-axis site. Of course, the platinum channel may be used in this detector as an additional upgrade.
\item[Second baseline]
A second baseline is already included as upgrade in the IDS-NF baseline setup. Of course, it requires an additional MIND, which we assume to have 50~kt. As baseline, typically $7500 \, \mathrm{km}$ is used for small $\theta_{13}$. However, the HENF (and also LENF) optimization for large $\theta_{13}$ might be different, and will be studied.
Note that if both baselines are operated simultaneously, fewer muons can be injected in each storage ring.
\item[Detector mass upgrade]
For small $\theta_{13}$, a detector mass upgrade of the MIND detector from 50~kt to 100~kt at the shorter baseline will be considered. Such an upgrade is currently discussed already within the IDS-NF. For large $\theta_{13}$, an upgrade of the TASD would disproportionally increase the detector cost compared to the whole complex.
\item[Knowledge of matter density profile] For large $\theta_{13}$, the matter density uncertainty is known to be one of the limiting factors (see, \eg, \Refs~\cite{Huber:2002mx,Ohlsson:2003ip,Huber:2006wb}), together with systematics. Therefore, we test the impact of the matter density knowledge. This implies that we also test how much the performance would improve (or how the optimization would change) in the presence of different matter density uncertainties.
\end{description}

\subsection{Simulation techniques}

The total running time of our experiments is assumed to be ten years, unless
stated otherwise. For the useful number of muon decays, we use the ``scale factor''
SF, where SF=1 corresponds to $2.5 \cdot 10^{20}$ useful muon decays per year,
which is the IDS-NF standard per baseline and polarity. In one storage ring,
both polarities ($\mu^+$ and $\mu^-$ decays) are operated simultaneously. If only one baseline
is operated, SF=2, because all muons are injected in the same storage
ring. The scale factor re-scales the total luminosity, and can be used to test the
impact of any re-scaling of statistics. Therefore, we sometimes use it as parameter.

For the sensitivity analyses we use the oscillation parameter values
(see, \eg, \Refs~\cite{GonzalezGarcia:2007ib,Schwetz:2008er}): $\sdm=7.65 \cdot 10^{-5} \,
\mathrm{eV}^2$, $|\ldm|=2.40 \cdot 10^{-5} \, \mathrm{eV}^2$, $\sin^2
\theta_{12}=0.304$, $\sin^2 \theta_{23}=0.500$, and a normal hierarchy, unless stated
otherwise. We impose external $1\sigma$ errors on $\sdm$ (4\%) and
$\theta_{12}$ (4\%) as conservative estimates for the current
measurement errors~\cite{Schwetz:2008er}. We do not include an external measurement of
the atmospheric parameters. In addition,
we include a 2\% matter density uncertainty, unless stated otherwise~\cite{Geller:2001ix,Ohlsson:2003ip}.
For the experiment simulation, we use the GLoBES software~\cite{Huber:2004ka,Huber:2007ji}.
All ``unused'' oscillation parameters are marginalized over, such as all parameters except for $\stheta$ for the $\stheta$ discovery reach.

\section{Neutrino factory for large $\boldsymbol{\theta_{13}}$ ($\boldsymbol{\theta_{13}}$ found)}
\label{sec:largetheta13}

In this section, we discuss the strategy for a neutrino factory if $\stheta$ is found by the next
generation of experiments, \ie, $\stheta \gtrsim 0.01$. We first of all discuss the minimal requirements for a neutrino factory in terms of baseline and luminosity without any additional information or upgrades. Then we study the baseline optimization including upgrades. Finally, we discuss two-baseline configurations for large $\theta_{13}$.

\subsection{Minimal neutrino factory}

Here we investigate the minimal requirements for a neutrino factory. We consider both the LENF and HENF,
but keep in mind that ``minimal'' clearly refers to the LENF. In addition, we only consider one baseline.
We assume that the measurement should be independent of that from other experiments, such as Daya Bay, and we do not consider any additions or upgrades.

Compared to the small $\stheta$ case, it is much easier to define a minimum wish list if $\stheta$ has already been observed. Here we follow the minimum wish list in \Ref~\cite{Winter:2008cn} (which was discussed there in context of the beta beam):
\begin{enumerate}
\item
 $5\sigma$ independent confirmation of $\stheta>0$ (for any $\deltacp$).
\item
 $3\sigma$ determination of the mass hierarchy (MH) for {\em any} (true) $\deltacp$.
\item
 $3\sigma$ establishment of CP violation (CPV) for a certain fraction (such as 80\%) of all (true) $\deltacp$.
\end{enumerate}
The only ``arbitrary'' in this list is the fraction of $\deltacp$ for which CPV should be discovered. A fraction of 80\% corresponds to Cabibbo-angle precision, which can be motivated in quark-lepton complementarity scenarios (see, \eg, \Refs~\cite{Winter:2007yi,Niehage:2008sg}). Alternatively, it corresponds to the precision of the CP phase in the quark sector is measured.
In this wish list, point~1 is typically easy for most of the parameter space; therefore, we do not show it explicitely anymore. Point~2 typically requires a certain minimum baseline. Point~3 requires sufficient luminosity and an appropriate baseline window close to the oscillation maximum. Compared to  \Ref~\cite{Winter:2008cn}, we simplify the analysis somewhat and show the results only for particular choices of (true) $\stheta$ (not ranges allowed by the next generation of experiments). Of course, the choice of $\stheta$ will be motivated by the results from preceding experiments. In addition, we show the normal hierarchy only. 

\begin{figure}[t!]
\begin{center}
\includegraphics[width=0.45\textwidth]{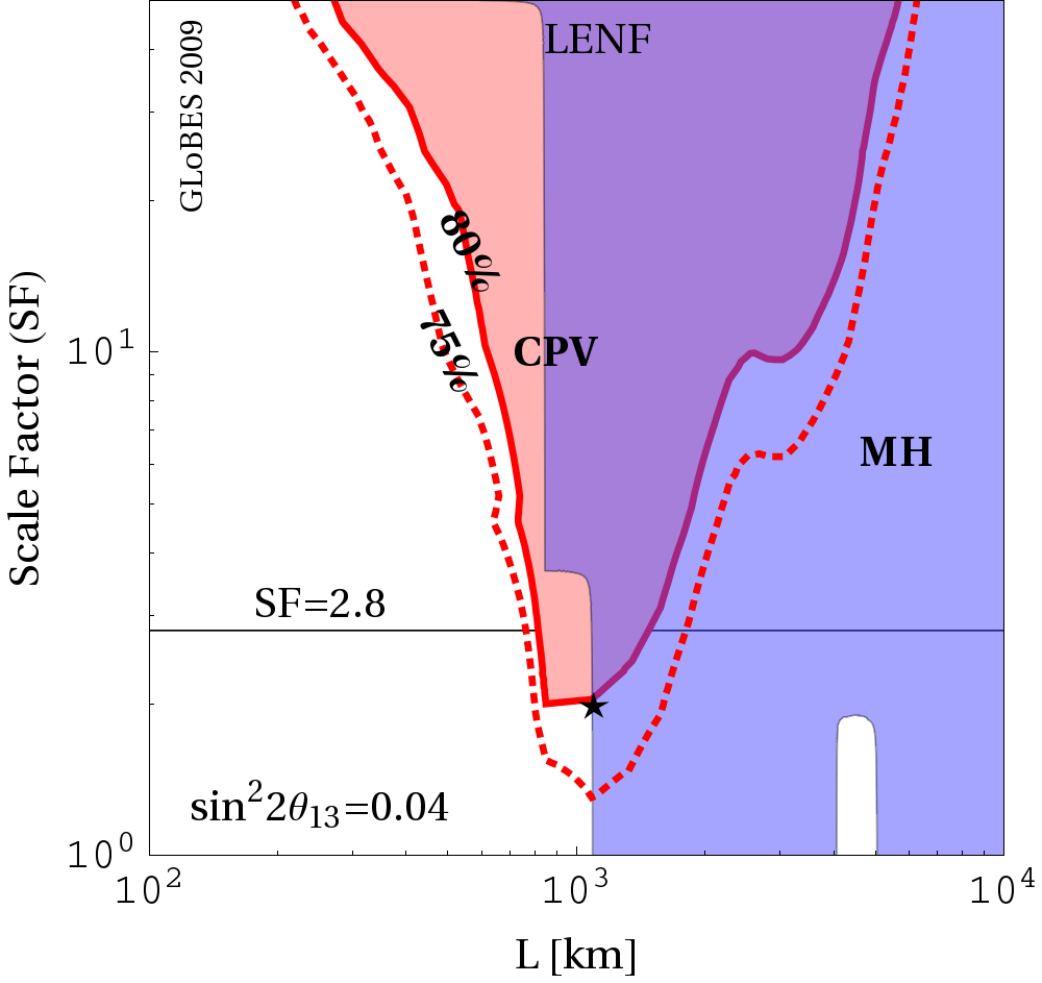} \hspace*{0.05\textwidth} \includegraphics[width=0.45\textwidth]{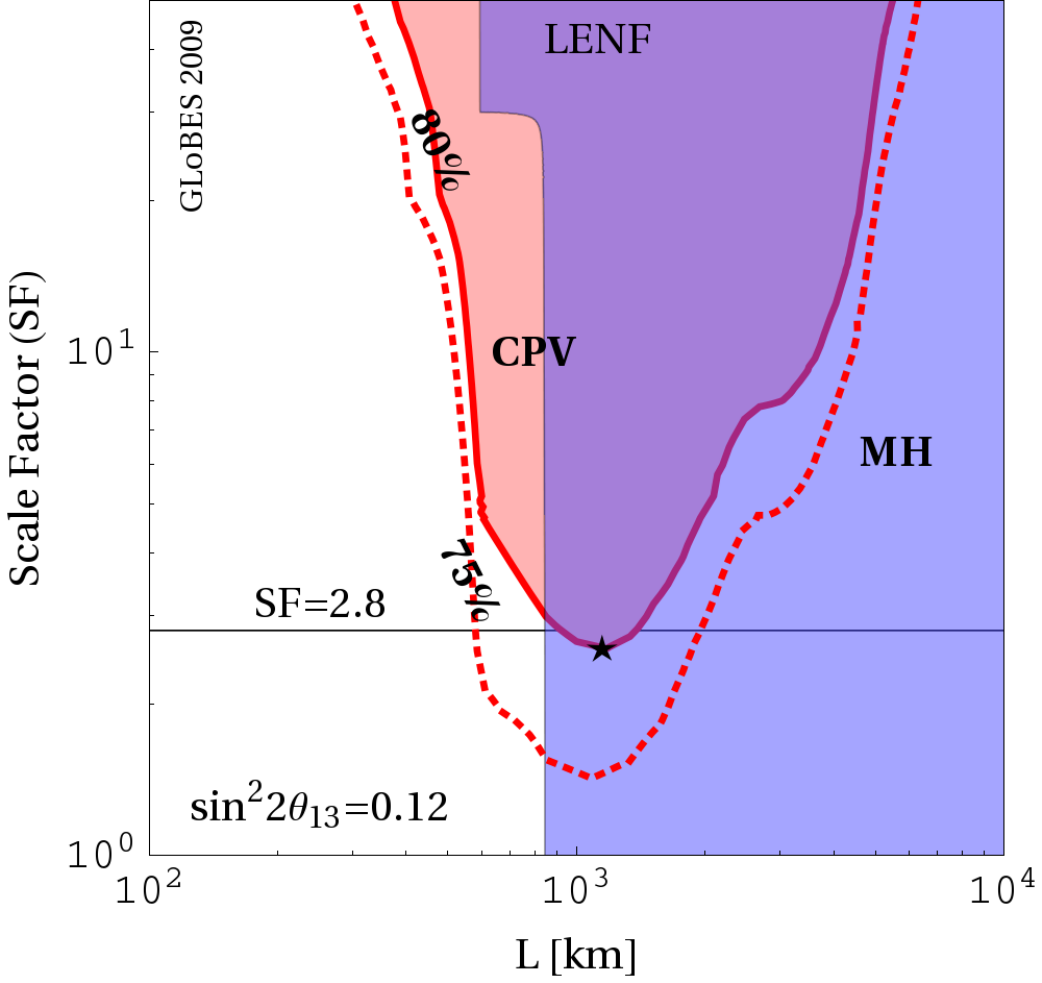}
\end{center}
\mycaption{\label{fig:mLENF}Discovery of  CPV (dark/red) and MH (medium gray/light blue) for the one baseline (minimal) LENF as a function of baseline and luminosity scale factor SF.  Discovery reach is given within the shaded regions at the $3\sigma$ CL, where for CPV a fraction of $\deltacp$ of 75\% or 80\% is required (as indicated), and for the MH a fraction of $\deltacp$ of 100\%. The stars show the baseline with the minimal SF: in the left panel $(1100 \, \mathrm{km},2.0)$ and in the right panel $(1150 \, \mathrm{km},2.6)$. The nominal luminosity is given by SF=2.8. Here the true value of $\sin^22\theta_{13}$ is chosen as given in the plot panels, and a normal hierarchy is assumed. The matter density uncertainty is assumed to be 2\%.}
\end{figure}

In order to identify the minimal version of the neutrino factory, we re-optimize the baseline, and, at the same time, identify the minimum luminosity for the optimal baseline with respect to the above performance indicators. 
For the LENF, we show in \figu{mLENF} the discovery reach for  CPV (dark/red) and MH (medium gray/light blue)  as a function of baseline and luminosity scale factor SF.  Discovery reach is given within the shaded regions at the $3\sigma$ CL, where for CPV a fraction of $\deltacp$ of 75\% or 80\% is required (as indicated), and for the MH a fraction of $\deltacp$ of 100\%. Sensitivity to both performance indicators is given in the overlap region, where one typically also has $\theta_{13}$ discovery potential for all $\deltacp$. The stars mark the points with the minimal SF where {\em all} performance indicators can be measured; they therefore show the ``minimal configurations''.
For the LENF, the minimal baseline is determined by the MH reach, and the minimal SF by the CPV reach. The nominal luminosity (SF=2.8) is sufficient for the CPV measurement for 80\% of all true $\deltacp$ and for the MH measurement for all $\deltacp$ in the baseline window $1100 \, \mathrm{km} \lesssim L \lesssim 1400 \, \mathrm{km}$ for both values of $\stheta$ (left and right panel). One can read off these figures that luminosity is clearly an issue for large $\stheta$. If, for instance, only a lower SF can be achieved, the CPV discovery reach decreases accordingly. The ``minimal'' (optimal) LENF, \ie, the one with the lowest SF, is in both panels at about $L \simeq 1100 \, \mathrm{km}$. However, the FNAL-DUSEL baseline $L = 1290 \, \mathrm{km}$ is close enough to optimum.

\begin{figure}[t!]
\begin{center}
\includegraphics[width=0.3\textwidth]{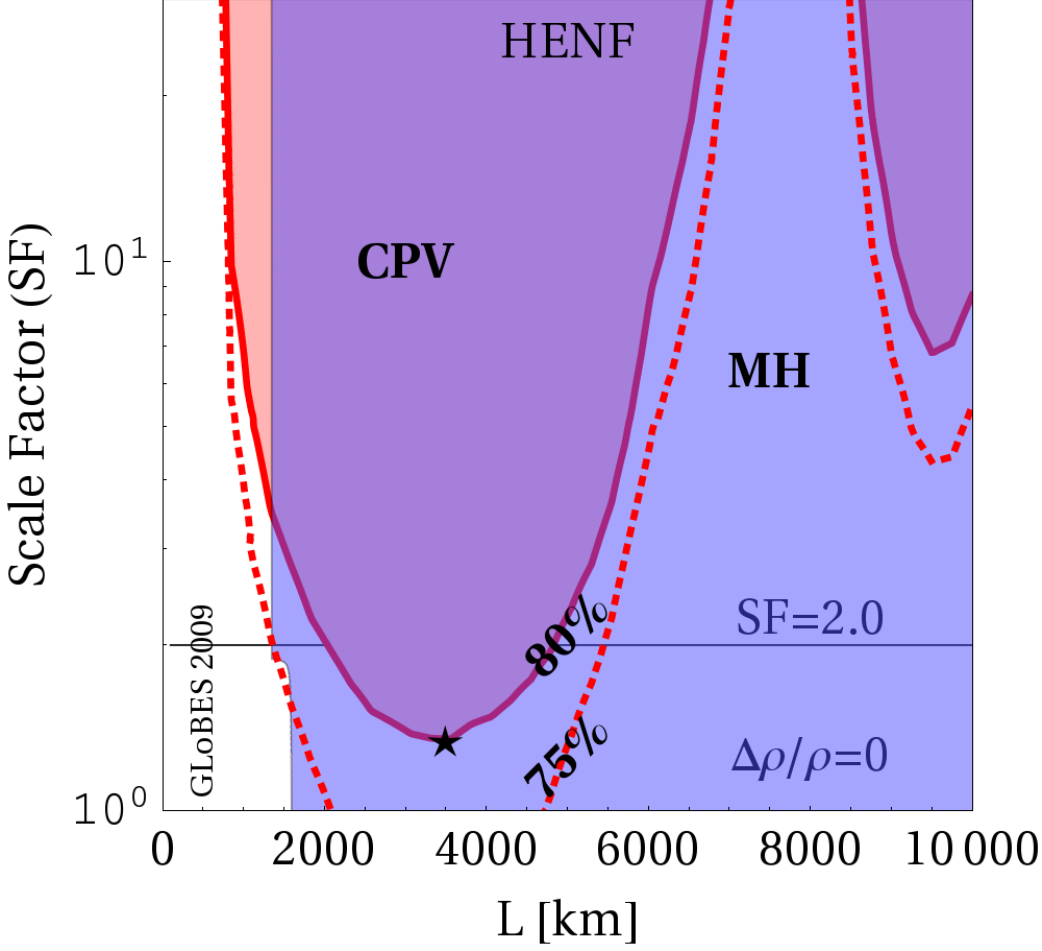}
\includegraphics[width=0.3\textwidth]{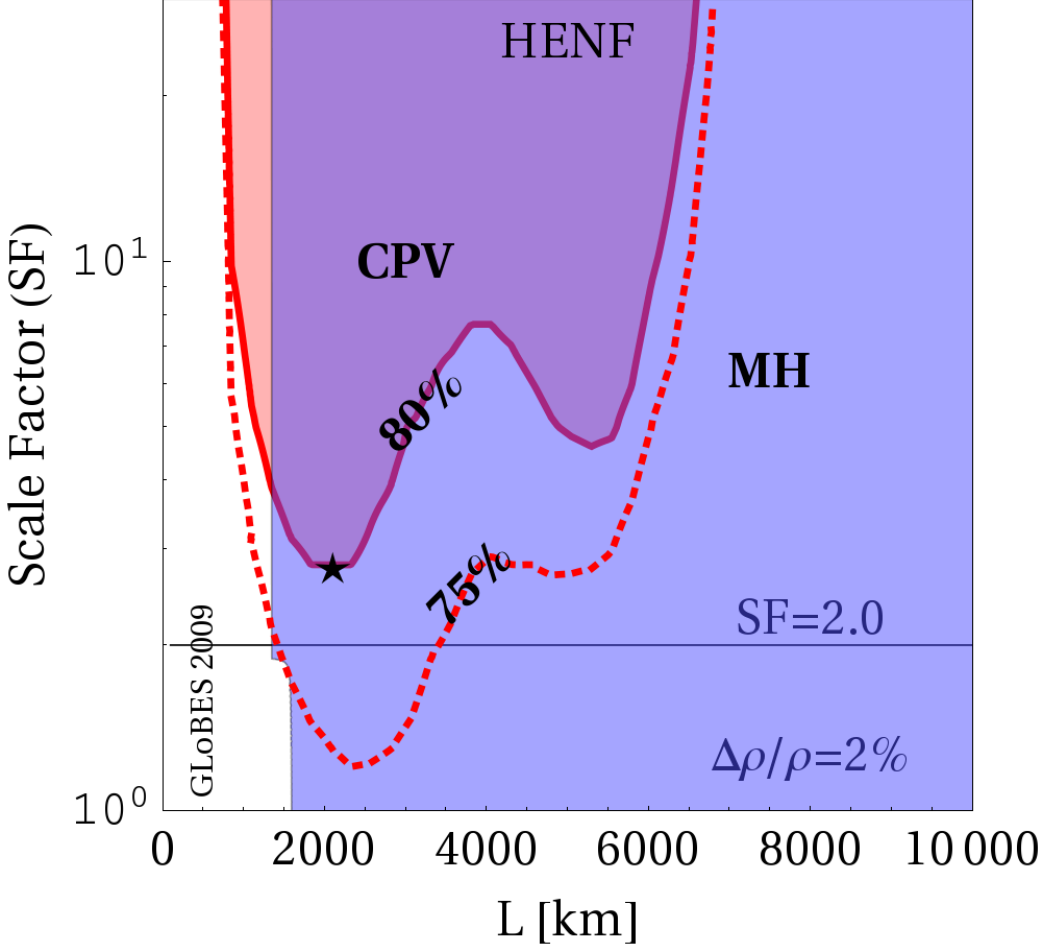} \includegraphics[width=0.3\textwidth]{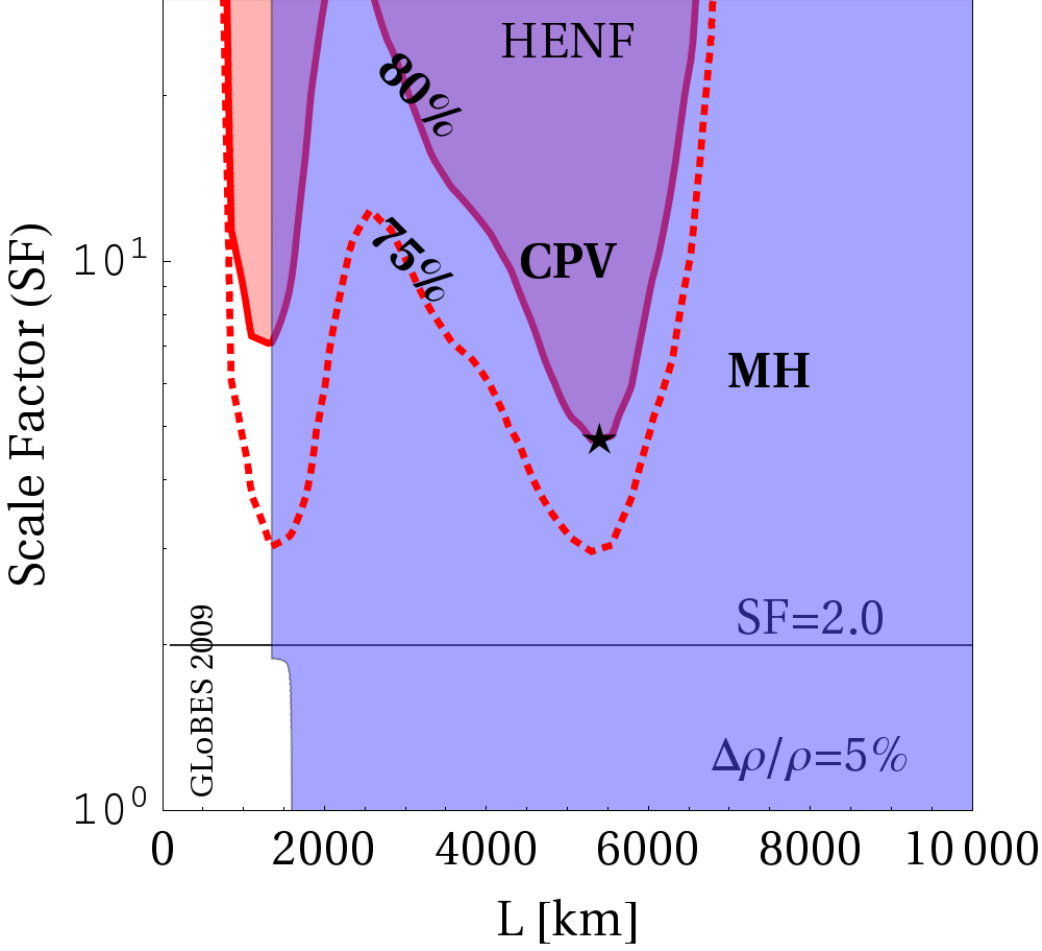}
\end{center}
\mycaption{\label{fig:mHENF} Discovery of  CPV (dark/red) and MH (medium gray/light blue) for the one baseline (minimal) HENF as a function of baseline and luminosity scale factor SF. Discovery reach is given within the shaded regions at the $3\sigma$ CL, where for CPV a fraction of $\deltacp$ of 75\% or 80\% is required (as indicated), and for the MH a fraction of $\deltacp$ of 100\%. The stars show the baseline with the minimal SF: in the left panel $(3500 \, \mathrm{km},1.35)$, in the middle panel $(2100 \, \mathrm{km},2.8)$, in the right panel $(5400 \, \mathrm{km},4.8)$. The nominal luminosity is given by SF=2. Here the true value of $\sin^22\theta_{13}=0.08$ and a normal hierarchy are assumed, and the matter density uncertainty is varied from the left to the right, as shown in the plot panels.}
\end{figure}

For the HENF, we show in \figu{mHENF} the discovery reach for CPV (dark/red) and MH (medium gray/light blue) for the one baseline (minimal) HENF as a function of baseline and luminosity scale factor SF. Here we show the dependence on the matter density uncertainty, because this is the main impact factor for the HENF, and instead choose only one value of (true) $\stheta=0.08$. Compared to the LENF, the CPV discovery reach typically determines the minimal configuration denoted by the stars. One can easily see that both performance and baseline optimization strongly depend on the degree the matter density profile is known. For instance, in the left panel, $\Delta \rho/\rho = 0$ (perfectly known matter density profile), and a fraction of $\deltacp$ of 80\% can be easily achieved in a baseline window
$2000 \, \mathrm{km} \lesssim L \lesssim 5000 \, \mathrm{km}$ for the nominal SF=2, where also the MH can be measured for all $\deltacp$. For more realistic matter density uncertainties, however, the situation becomes more complicated (\cf, middle and right panels for 2\% and 5\%, respectively). The CPV discovery reach, in these cases, exactly deteriorates in the window which is optimal for no matter density uncertainty, and two local minima, one at a shorter baseline $L \simeq 2000 \, \mathrm{km}$ and one at a longer baseline $L \simeq 5500 \, \mathrm{km}$, remain (\cf, ``$\pi$-transit'' problem~\cite{Huber:2002mx}). The absolute reach is in both cases above the threshold of 80\%. Therefore, a reliable optimization of the HENF without upgrades for large $\stheta$ is only possible if the matter density profile is precisely known.

Since for realistic matter density uncertainties the LENF outperforms the HENF at the nominal luminosity and the HENF relies on a higher effort on the accelerator side, the LENF is probably the best ``minimal'' version of a neutrino factory for large $\stheta$. However, it may be upgraded for different purposes later. For instance, the search for new physics, such as non-standard interactions, requires higher muon energies~\cite{Kopp:2008ds}. Therefore, we nevertheless consider the HENF for large $\stheta$, especially in the context of upgrades.

\subsection{Single baseline neutrino factory: Upgrades and optimization }

\begin{figure}[t!]
\begin{center}
\includegraphics[width=0.45\textwidth]{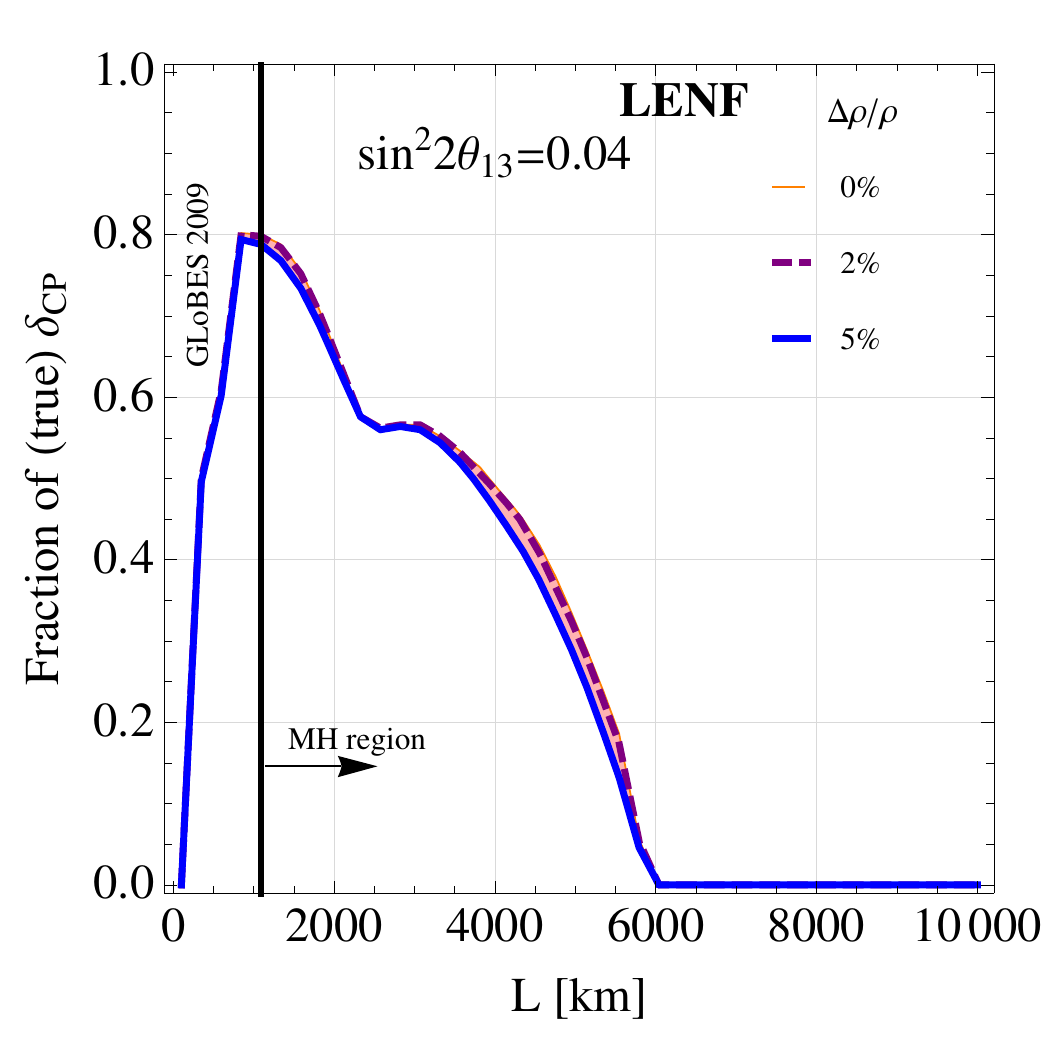} \hspace*{0.05\textwidth} \includegraphics[width=0.45\textwidth]{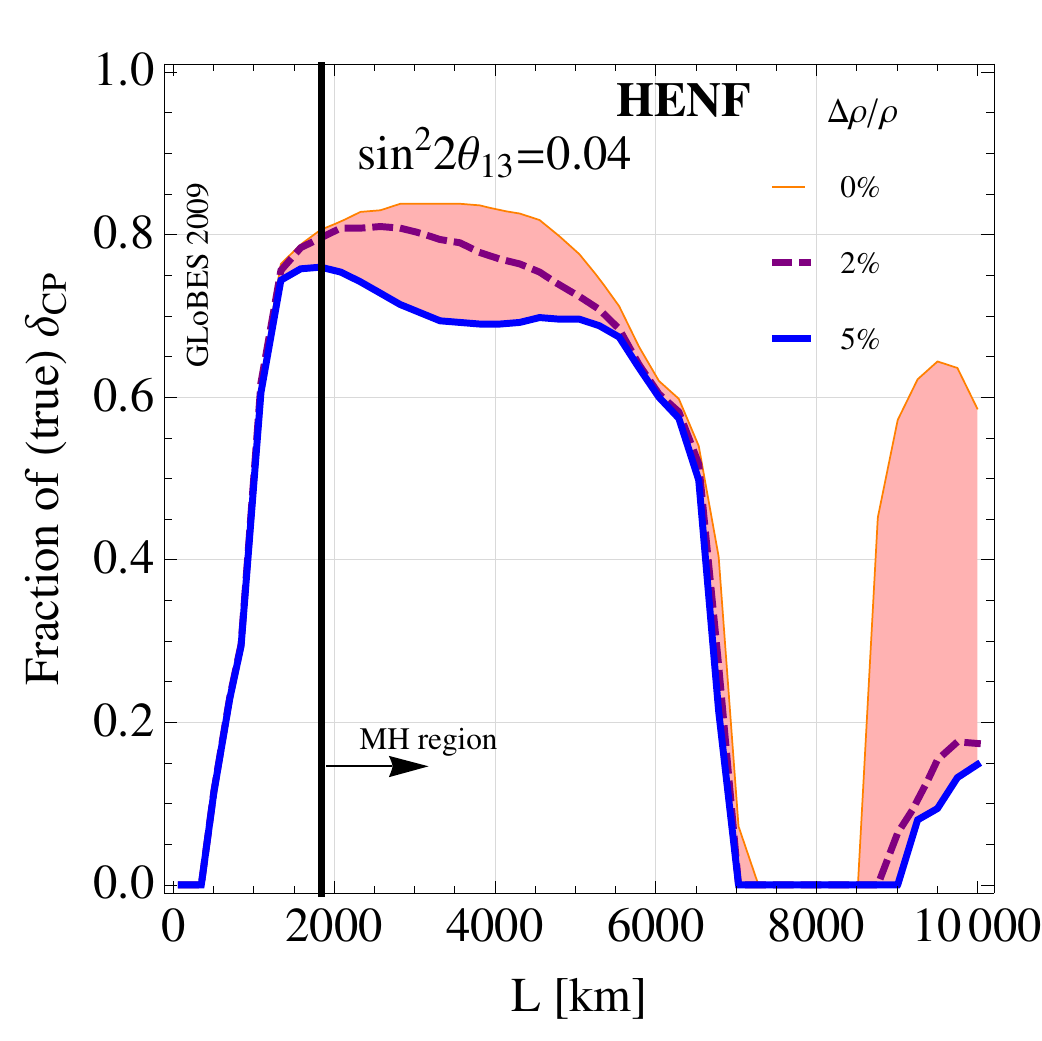}
\end{center}
\mycaption{\label{fig:optNF-CPV} Fraction of $\deltacp$ for which CPV will be discovered as a function of baseline $L$ for the LENF (left panel) and HENF (right panel); $3 \sigma$ CL. The different curves show the impact of the matter density uncertainty, as given in the plot legends.  The vertical lines mark the regions where also the MH can be determined for all $\deltacp$ at the $3 \sigma$ CL. Here $\stheta=0.04$, SF=2, and a normal hierarchy is assumed.}
\end{figure}

Here we discuss the LENF and HENF baseline optimization for large $\theta_{13}$ together with possible upgrades or external input. In this case, we choose the nominal SF=2 for both LENF and HENF. This choice has the advantage that LENF can also be interpreted as HENF-OAD.

As the first aspect, we show in \figu{optNF-CPV} the fraction of $\deltacp$ for which CPV will be discovered as a function of baseline $L$ for the LENF (left panel) and HENF (right panel). The different curves show the impact of the matter density uncertainty, as given in the plot legends. 
For the LENF (left panel), we find that the fraction of $\deltacp$ drops sharply off the optimal baseline window identified above. The matter density uncertainty hardly has any impact on the absolute performance or optimization. A fraction of $\deltacp$ of 80\% can be reached at about the peak without any further upgrades. Note again that the LENF could also be HENF-OAD (therefore we chose SF=2 consistently), and option which we will discuss below.
For the HENF (right panel), the optimal baseline depends on the matter density uncertainty, and also somewhat on $\stheta$ (here we only show one example). The fraction of $\deltacp$ of 80\% can (for $\stheta=0.04$) be reached if the matter density is small enough  ($\lesssim 2\%$). For the chosen value of $\stheta$, the optimal baseline for both CPV and MH is around $2 000 \, \mathrm{km}$ to $4 000 \, \mathrm{km}$ (depending in matter density uncertainty and $\stheta$). Again, the LENF is the more robust version for large $\stheta$, because the optimization does not depend on the matter density uncertainty.

\begin{figure}[t!]
\begin{center}
\includegraphics[width=0.45\textwidth]{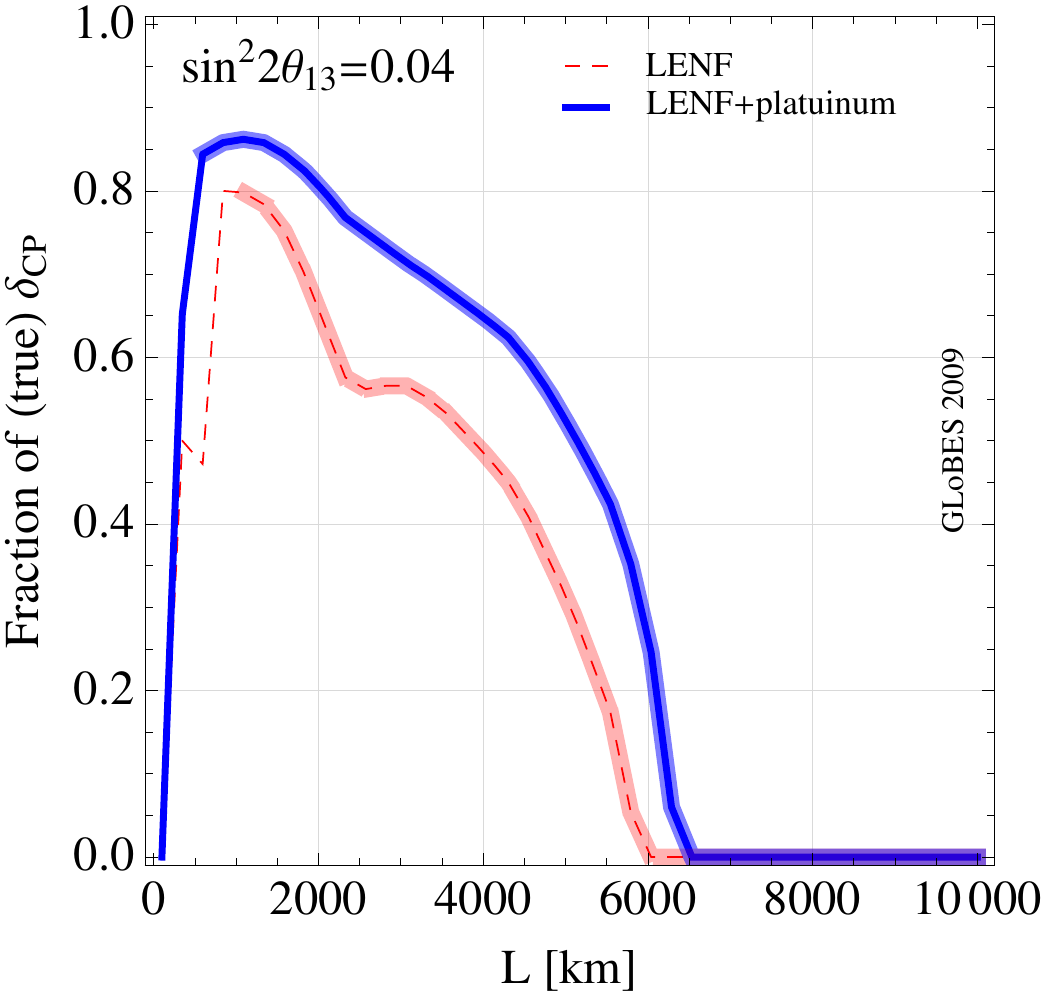} \hspace*{0.05\textwidth} \includegraphics[width=0.45\textwidth]{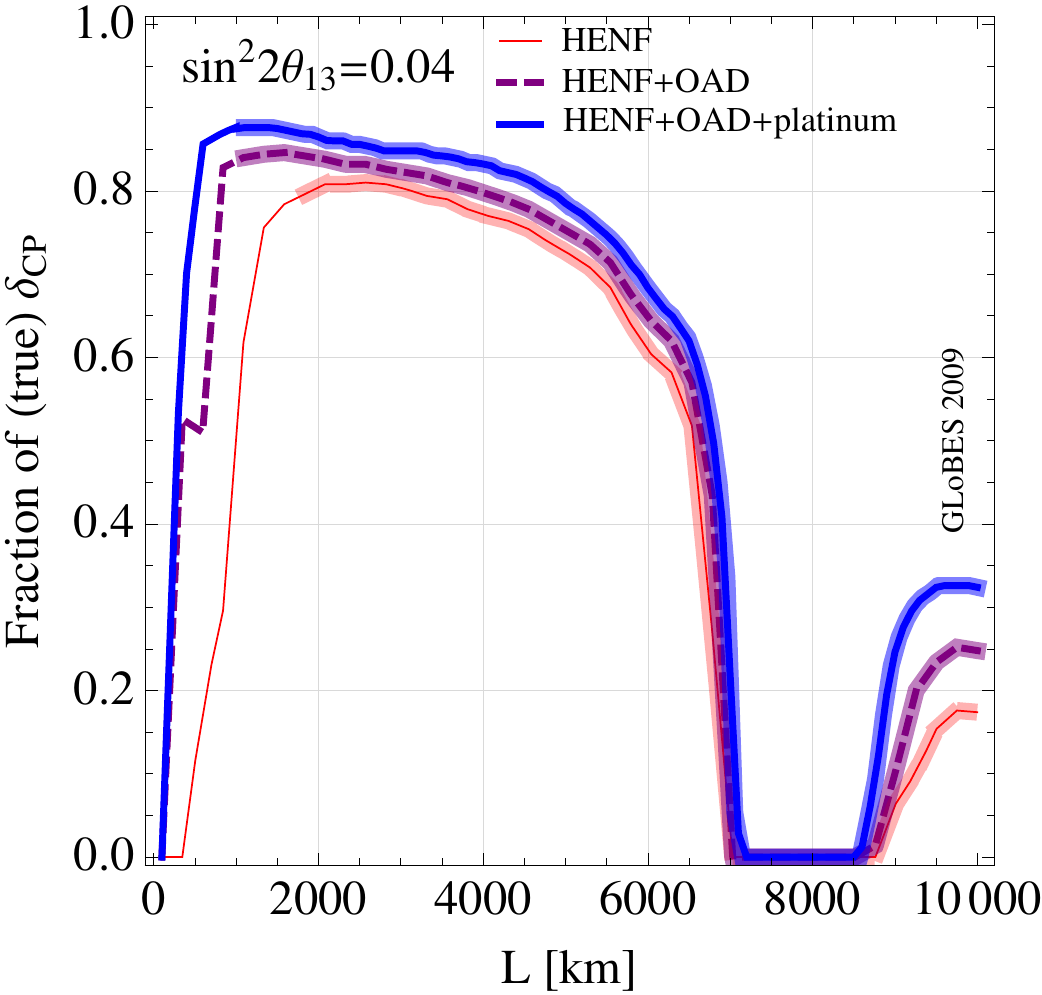}
\end{center}
\mycaption{\label{fig:optNF}  Fraction of $\deltacp$ for which CPV will be discovered as a function of baseline $L$ for the LENF (left panel) and HENF (right panel); $3 \sigma$ CL. The different curves show different upgrade options: off-axis TASD (OAD) and platinum channel. The thick curves give the baseline range where also the mass hierarchy can be determined for {\em any} $\deltacp$ at the $3 \sigma$ confidence level. Here $\stheta=0.04$, SF=2, a normal hierarchy and a $2\%$ of the matter density uncertainty are assumed.}
\end{figure}

As far as possible upgrades are concerned, we show in \figu{optNF} the fraction of $\deltacp$ for which CPV will be discovered as a function of baseline $L$ for the LENF (left panel) and HENF (right panel), where we have chosen a matter density uncertainty of 2\%. The thick curves give the baseline range where also the mass hierarchy can be determined for {\em any} $\deltacp$ at the $3 \sigma$ confidence level. 
For the LENF (left panel), the most plausible upgrade may be the platinum channel, which clearly increases the absolute performance for large $\stheta$ (left panel) up to a fraction of $\deltacp$ 85\%. In addition, the optimal baseline window becomes larger, and a short baseline, such as FNAL-Soudan, could be sufficient to break the mass hierarchy degeneracy (where the short baseline cutoff in the CPV discovery reach comes from).

A different possibility together with upgrades my be the HENF (right panel). An additional OAD would combine the virtues of the LENF and HENF. In this case, the optimal baseline becomes shorter, possibly as short as FNAL-DUSEL. Together with the platinum channel (in the OAD), CPV can be even measured for almost 90\% of all $\deltacp$ at a baseline of about $1100 \, \mathrm{km}$. At this baseline, the matter density uncertainty hardly has any impact. Note, however, that even longer baselines up to $4500 \, \mathrm{km}$ still lead to a CPV discovery for more than 80\% of all $\deltacp$. In this case, the matter density profile has to be controlled at the level of a few percent. The correlation between the profiles seen by the off- and on-axis detectors (distance only about $38 \, \mathrm{km}$ at $L=4000 \, \mathrm{km}$), which we have included in the figure, helps somewhat.
Therefore, if for some reason (such as new physics searches) a HENF may be the preferred option, the combination with the OAD-TASD could be the most interesting upgrade. It allows for a relatively wide baseline window.

We have also tested the impact of 5~year Daya Bay data, simulated in \Ref~\cite{Huber:2009cw}, which may affect the optimization for large $\theta_{13}$. However, we have not found any significant impact, neither on absolute performance, nor on the baseline optimization, by the combination with the Daya Bay data. However, note that a certain simulated value of $\stheta$ is always assumed for the optimization, which is only known within the range expected from the next generation of experiments.

\subsection{Double baseline neutrino factory}

\begin{figure}[t!]
\begin{center}
\includegraphics[width=0.285\textwidth]{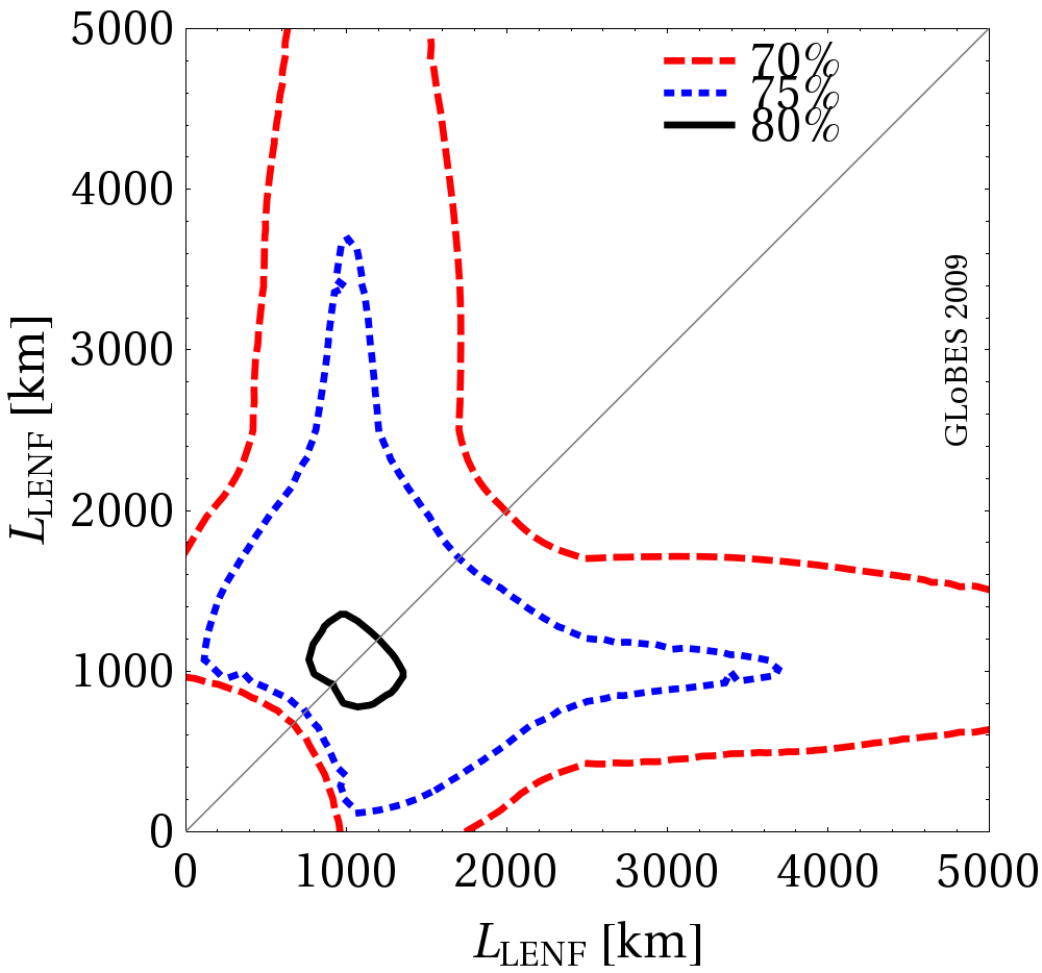}
\includegraphics[width=0.3\textwidth]{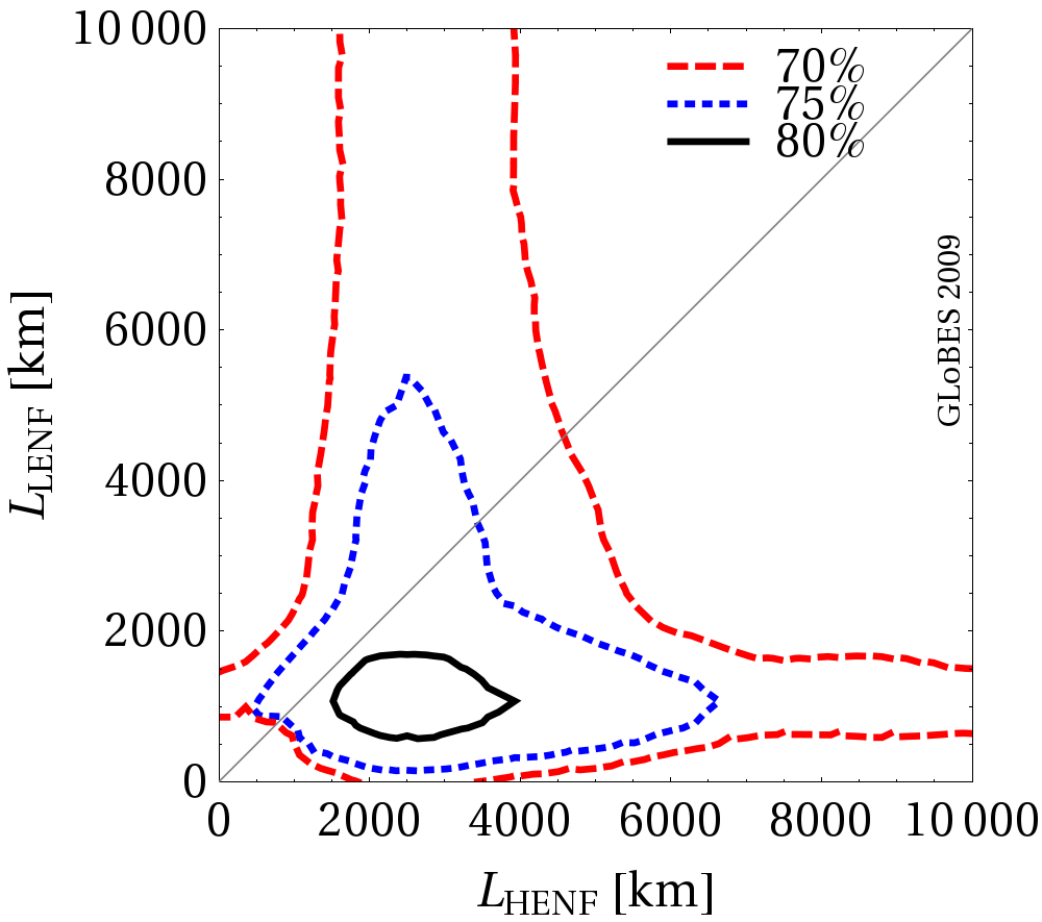} \includegraphics[width=0.3\textwidth]{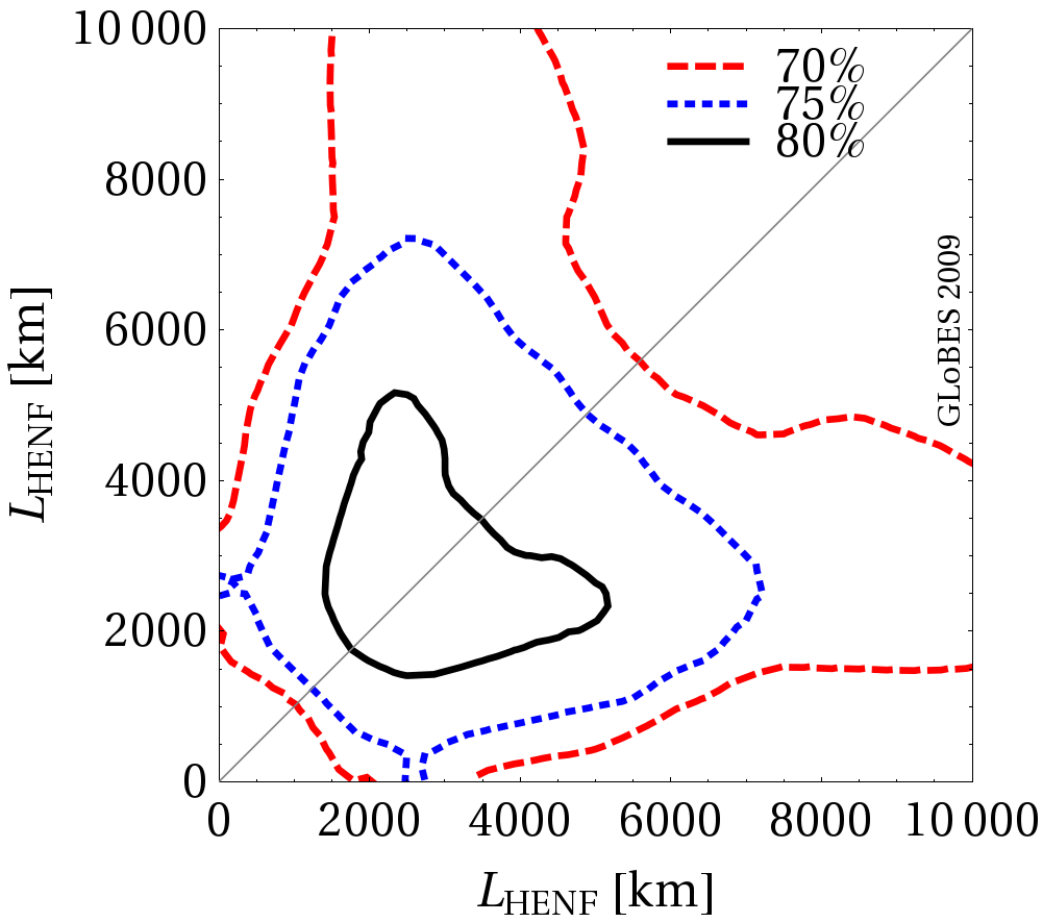}
\end{center}
\mycaption{\label{fig:LoptNF} Two baseline optimization for CPV: Fraction of $\deltacp$ for which CPV is discovered as a function of the two neutrino factory baselines; $3 \sigma$ CL. Here the combinations LENF-LENF (left panel), HENF-LENF (middle panel), and HENF-HENF (right panel) are shown. Note that LENF can also refer to the OAD in this figure.  Here a 2\% matter density uncertainty, SF=1 (each baseline), the true value $\sin^22\theta_{13}=0.08$, and a normal hierarchy are assumed. Here no upgrades are included.}
\end{figure}

So far, we have assumed that only one baseline is used. A double baseline neutrino factory could also be an option for large $\theta_{13}$, where the optimization might be different from the small $\theta_{13}$ case studied in the literature~\cite{Huber:2003ak,Kopp:2008ds}. A double baseline neutrino factory could be interesting in terms of the following upgrades:
\begin{description}
\item[Energy upgrade:]
 LENF, followed by a HENF at a (possibly) different baseline
\item[Baseline upgrade:]
 LENF at two different baselines or HENF at two different baselines
\item[Additional off-axis detector:]
 HENF, followed by an OAD at a (possibly) different baseline
\end{description}

We study these three options in \figu{LoptNF} in terms of the fraction of $\deltacp$ for which CPV is discovered as a function of two baselines. For the baseline combinations, we have LENF-LENF, HENF-LENF (or OAD), HENF-HENF, where we choose SF=1 for each baseline. This means that we assume that two storage rings are required, and the muons are equally shared among the rings. In principle, LENF-LENF or HENF-HENF for the same baseline (along the diagonal) correspond to the previously studied cases. However, note that the external input (such as the knowledge on the solar parameters) is added twice here, and that the matter density uncertainty is assumed to be uncorrelated between the two baselines. Therefore, they cannot be directly compared to the previous results.

The main conclusion from \figu{LoptNF} is that in all cases the LENF and HENF baseline optima are obtained almost uncorrelated with each other.  This means that, if one combines two baselines and allows for two different storage rings, the LENF is optimal at about $L=1000 \, \mathrm{km}$, the HENF at about $L=2000 \, \mathrm{km}$ to $4000 \, \mathrm{km}$, which roughly corresponds to the individual one baseline optimizations.\footnote{Only in some corners of the parameter space, such as HENF+HENF for $\stheta=0.12$ and 5\% matter density uncertainty, we have found that combining a shorter with a longer baseline may somewhat help.}
Therefore, for any energy or baseline upgrade, one would follow the strategy from the individual baseline optimizations. For the combinations LENF+LENF and HENF+HENF, the most plausible upgrade may then be a detector mass upgrade.

\vspace*{0.5cm}

In summary, a LENF at a baseline of about 1100 to 1400~km may be the most plausible neutrino factory option for large $\stheta$. If possible, it should rely on electron neutrino appearance (platinum channel) as well. We pointed out that if new physics searches indicate that a HENF may be desirable, the LENF detector could also be used as OAD for the HENF, with significant impact on the sensitivity. 

\section{Staging for small $\boldsymbol{\theta_{13}}$ ($\boldsymbol{\theta_{13}}$ not found)}
\label{sec:smalltheta13}

After several years of data taking from Daya Bay and the other next generation experiments, we will know whether $\stheta \lesssim 0.01$~\cite{Huber:2009cw}. For this part, we therefore assume that $\theta_{13}$ has not been found by these experiments, which corresponds to  $\stheta \lesssim 0.01$. In  this case, the combination with Daya Bay will not help very much. The same applies to the platinum channel, which is limited by the charge identification capabilites. How can one then build a neutrino factory step by step, while taking into account the knowledge from the preceding phases?

What we already know is that $L \simeq 4000 \, \mathrm{km}$ is optimal for CPV~\cite{Huber:2006wb},  $L \simeq  7500 \, \mathrm{km}$ for a risk-minimized mass hierarchy measurement~\cite{Huber:2006wb} and an excellent $\theta_{13}$ sensitivity~\cite{Huber:2003ak} (which approximately corresponds to the worst case discovery reach). The combination is good for all performance indicators~\cite{Kopp:2008ds}. However: These are optimized in the $\theta_{13}$ direction, \ie, discovery for as small as possible $\theta_{13}$. If we know, at some point, how large $\theta_{13}$ is, the discussion may change, depending on the priorities we have. 

In this section, we therefore sketch a plausible staging scenario, starting with a low energy neutrino factory.
As upgrade options, we consider increasing the muon energy, adding another baseline, and increasing the detector mass. For the sake of simplicity, we assume that the accelerator complex delivers $10^{21}$ useful muon decays for both polarities per year. If there are two storage rings operated with the same muon energy, these muons can be (at least in principle) split arbitrarily between the rings. We proceed in three phases of data taking, five years each. Of course, there may be idle times between these phases, such as because of construction. Phase~I represents a low energy neutrino factory, phase~II includes the energy upgrade to a high energy neutrino factory, and phase~III considers additional upgrades, such as a larger detector or an additional baseline. In any phase, we combine the data with the preceding phase.

\subsection{Phase~I: Low energy neutrino factory}
 
\begin{figure}[t]
\begin{center}
\includegraphics[width=0.48\textwidth]{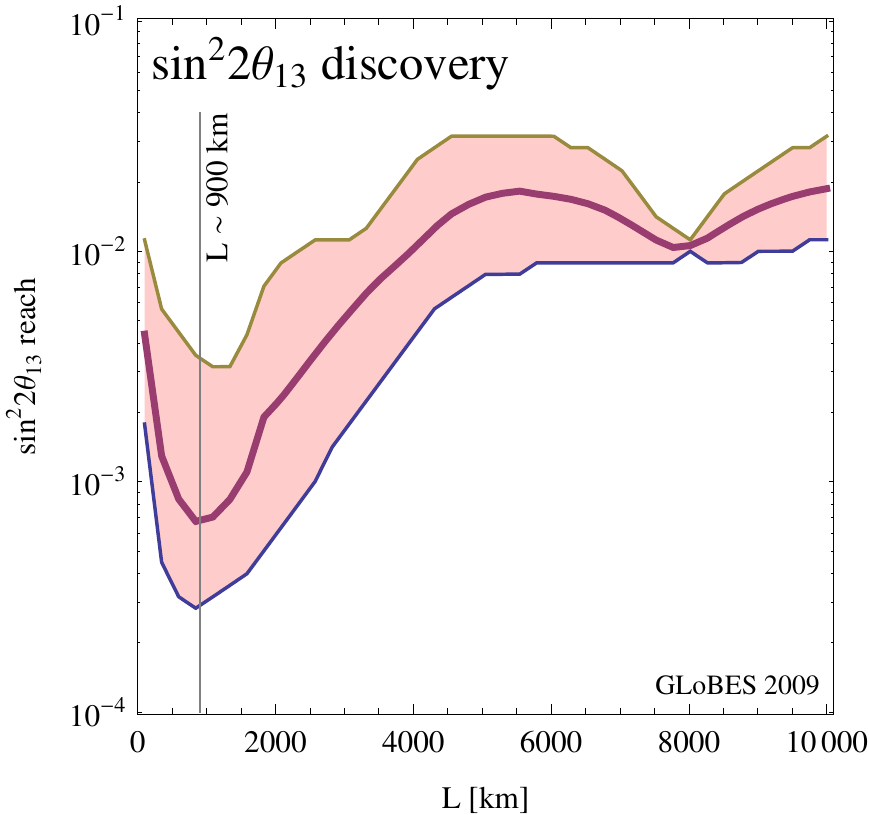} \hspace*{0.05\textwidth} \includegraphics[width=0.45\textwidth]{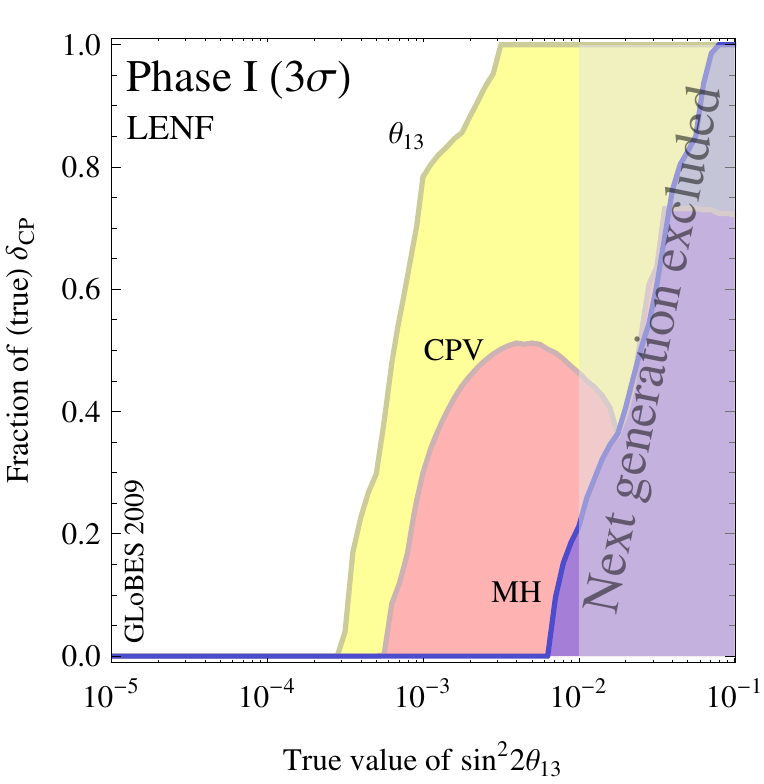}
\end{center}
\mycaption{\label{fig:resultP1} Phase~I: $\theta$ discovery reach ($3 \sigma$) as a function of baseline (left panel) and
all discovery reaches ($3 \sigma$) as a function of $\stheta$ and fraction of $\deltacp$ for the optimal $\theta_{13}$ baseline $L=900 \, \mathrm{km}$ (right panel). In the left panel, the best case (lower curve), worst case (upper curve) and ``typical'' (thick curve) true $\deltacp$ is shown. The ``typical'' $\deltacp$ corresponds to the median, \ie, the performance is better for 50\% of all $\deltacp$ and worse for 50\%. Here a normal hierarchy is assumed.}
\end{figure}

If $\theta_{13}$ is not discovered by the next generation of experiments, we assume that in phase~I of a neutrino factory a low energy version with $E_\mu=4.12 \, \mathrm{GeV}$ and a magnetized TASD as detector is operated. The main priority will be the search for $\theta_{13}$, which means that the machine should be optimized for that. Our phase~I has five years of operation with $10^{21}$ useful muon decays per year in both polarities, \ie, SF=2. We show in \figu{resultP1}, left panel, the baseline optimization for the $\theta_{13}$ discovery reach.  In this figure, the best case (lower curve), worst case (upper curve) and ``typical'' (thick curve) true $\deltacp$ is shown. The ``typical'' $\deltacp$ corresponds to the median, \ie, the performance is better for 50\% of all $\deltacp$ and worse for 50\%. Obviously, a baseline of about $900 \, \mathrm{km}$ is close-to-optimal for the typical $\deltacp$ (thick curve), as it was for for CP violation for large $\theta_{13}$ in \Ref~\cite{Huber:2007uj}. For a risk-minimized performance (upper curve), somewhat longer baselines are preferred. In summary, baselines between about  $500 \, \mathrm{km}$ and $1500 \, \mathrm{km}$ are sufficiently good. We use, in the following, $L=900 \, \mathrm{km}$, which will allow for a $\stheta$ discovery for values between 0.003 and 0.0003 for the normal hierarchy, depending on the true value of $\deltacp$.

In \figu{resultP1}, right panel, we show the mass hierarchy, CP violation and $\theta_{13}$ discovery reaches of phase~I as a function of $\theta_{13}$ and $\deltacp$. The region, which is accessible to the next generation of experiments and covered by the previous section, is shaded in gray. One can easily read off this figure that CP violation is only accessible for a small fraction of the parameter space if $\theta_{13}$ is discovered in phase~I, and the mass hierarchy can practically not be determined for $\stheta \lesssim 0.01$. Therefore, even in the case of a $\theta_{13}$ discovery, phase~II will be most likely needed. In case of a $\stheta$ discovery, we list some obtainable precisions for $\stheta$ for different values here (90\% CL, computed for  the true $\deltacp=\pi/2$):
\begin{eqnarray}
\text{For } \stheta=0.001: & \quad & 0.00048 \lesssim \stheta \lesssim 0.0037 \label{equ:1} \\
\text{For } \stheta=0.005: & \quad & 0.0037 \lesssim \stheta \lesssim 0.0072 \label{equ:2} \\
\text{For } \stheta=0.01: & \quad & 0.0081  \lesssim \stheta \lesssim 0.013 \label{equ:3} 
\end{eqnarray}

\subsection{Phase~II: Energy upgrade?}

\begin{figure}[t]
\begin{center}
\includegraphics[width=0.45\textwidth]{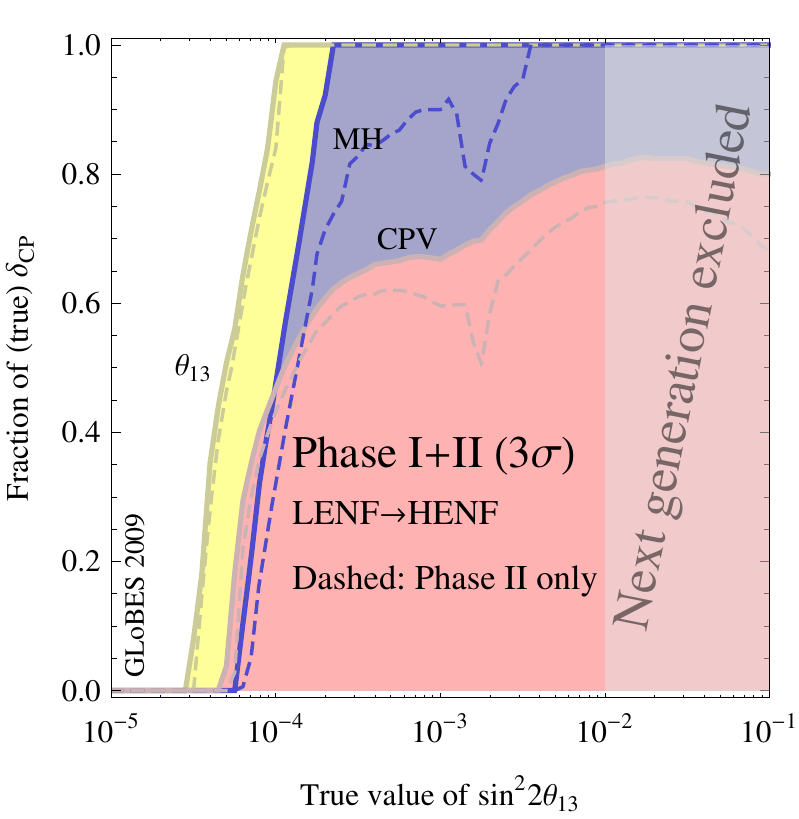} 
\end{center}
\mycaption{\label{fig:resultP2} Results from phases I+II. The different regions show the $\theta_{13}$, MH, and CPV discovery reaches ($3 \sigma$) as a function of $\stheta$ and fraction of $\deltacp$. The dashed curves represent phase~II alone.  Here a normal hierarchy is assumed.}
\end{figure}

If $\theta_{13}$ is not discovered in phase~I, the most plausible upgrade might be an energy upgrade, which comes together with a new detector and a new baseline (including storage ring) for optimal sensitivities.\footnote{See, \eg, \Ref~\cite{Huber:2006wb}: At least for small $\theta_{13}$, $L=900 \, \mathrm{km}$ is far away from optimal for the higher muon energy. Therefore, we do not consider placing the new detector at the $L=900 \, \mathrm{km}$ baseline.} 
 For the energy, we use the IDS-NF standard muon energy $E_\mu=25 \, \mathrm{GeV}$. It is typically discussed together with the MIND detector. We assume that phase~II corresponds to five years of data taking again, and we combine the data with phase~I. Note that all useful muons in phase~II are used for the new baseline (SF=2), \ie, $10^{21}$ useful muon decays per year in both polarities.

\subsubsection{$\boldsymbol{\theta_{13}}$ not discovered in phase~I}
 
  The optimization for small $\theta_{13}$ has been extensively studied in the literature, see \eg\ Ref~\cite{Huber:2006wb}. We follow the conventional strategy to aim for discovery reaches for as small as possible $\theta_{13}$:  L $\sim$ 4000~km is a good choice for CPV and also has an excellent $\theta_{13}$ discovery reach for most values of $\deltacp$. 

We show in \figu{resultP2} the result for this upgrade. With the upgrade, $\theta_{13}$ will be discovered for $\stheta$ as small as $10^{-4}$ for any value of $\deltacp$. In the case of a discovery, CPV and the mass hierarchy can be determined for a large fraction of the parameter space for $\stheta \gtrsim 10^{-4}$, beyond the next generation of experiments.
If the MH or CPV are not found, we discuss in \Sec~\ref{sec:p3nd} further possible upgrades.

Very interestingly, the combination with phase~I helps to resolve the degeneracies. This is illustrated by the dashed curves, which are computed for phase~II alone. Without phase~I, the degenerate solutions severely affect the CPV and MH discovery reaches, whereas the phase~I has just enough statistics to partially resolve the degeneracies.  In this case, the small matter effects in phase~I (because of the shorter baseline and lower energy) become an advantage, because the mass hierarchy degeneracy has a very different location.
In fact, we have also tested using double luminosity in phase~II alone, to check if there is a real synergy beyond the addition of statistics (\cf, \Ref~\cite{Huber:2002rs} for a more detailed discussion). In fact, we could identify such a synergy for the MH and for CPV for a part of the parameter space (the larger $\stheta$ values).

\subsubsection{$\boldsymbol{\theta_{13}}$ discovered in phase~I}

\begin{figure}[t]
\begin{center}
\includegraphics[width=\textwidth]{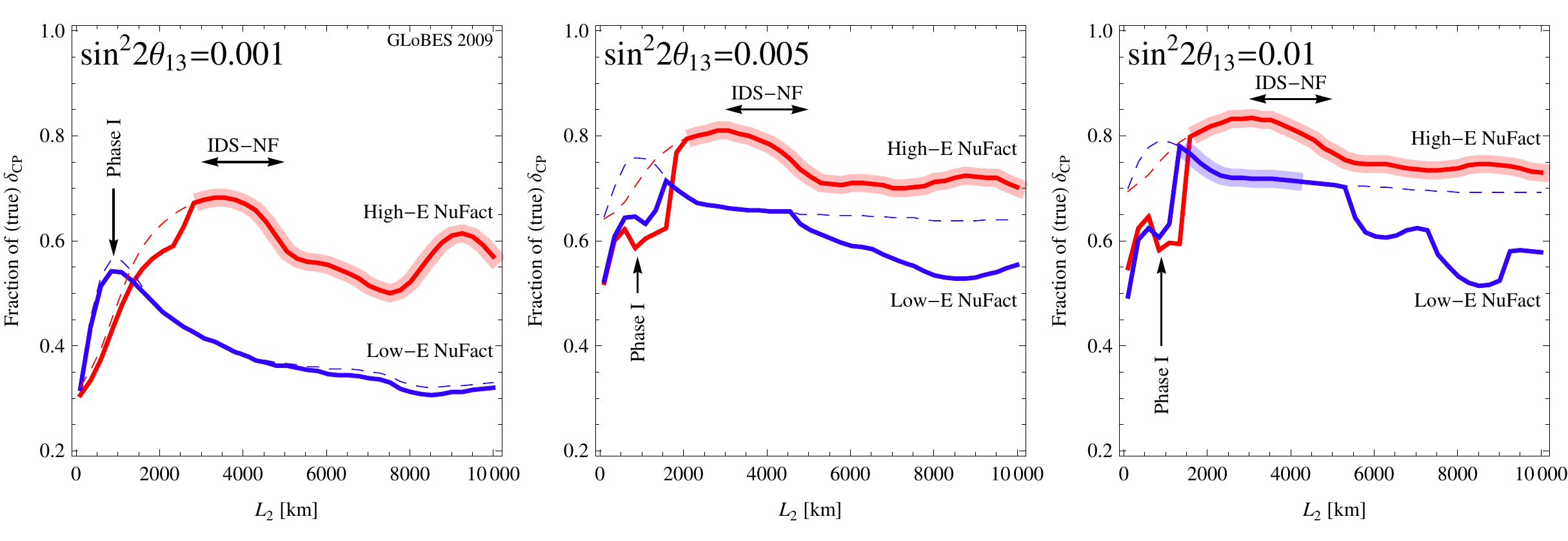}
\end{center}
\mycaption{\label{fig:base2} Baseline optimization for the phase~II baseline for the combination of phase~I and phase~II. Here the fraction of (true) $\deltacp$ for CPV discovery ($3\sigma$) is optimized for. The blue (dark) curves are drawn for a possible LENF in phase~II, the red (light) curves for a HENF (energy upgrade) in phase~II. The dashed curves do not include degeneracies. The thick curves show the baseline ranges where also the mass hierarchy can be determined for {\em any} $\deltacp$ at the $3 \sigma$ confidence level. The baseline for phase~I and the baseline range of the IDS-NF (for the shorter) baseline are marked. The different panels correspond to different values of $\stheta$, as they may be discovered in phase~I. Normal hierarchy assumed.}
\end{figure}

In this case, we re-consider the baseline optimization, depending on the $\theta_{13}$ found in phase~I.
For the sake of completeness, we study the physics of a second baseline with and without an energy upgrade, \ie, LENF or HENF in phase~II. The result for the optimization of the second baseline is shown in \figu{base2} for different best-fit values of $\stheta$, as they might be obtained in phase~I; \cf, Eqs.~(\ref{equ:1}) to (\ref{equ:3}). The plots show the fraction of (true) $\deltacp$ for CPV discovery ($3\sigma$) as a function of baseline. The blue (dark) curves are drawn for the LENF, the red (light) curves for the HENF (in phase~II). The dashed curves do not include degeneracies. The thick curves show the baseline range where also the mass hierarchy can be determined for {\em any} $\deltacp$ at the $3 \sigma$ confidence level. Note that the matter density uncertainty is assumed uncorrelated with the first baseline. 

For small values of $\stheta$ at the lower end of the range reachable in phase~I (left panel), the mass hierarchy cannot be measured for all $\deltacp$ with the LENF in phase~II, and the energy upgrade performs much better for CPV. The baseline choice $L_2 \simeq 4000 \, \mathrm{km}$ is precisely at the optimum for the HENF. Therefore, the strategy will be the same in the case of $\stheta$ not discovered in phase~I: energy upgrade with a MIND at $L_2 \simeq 4000 \, \mathrm{km}$. Note that the fraction of $\deltacp$ for LENF peaks at about $900 \, \mathrm{km}$, as it was optimized before.

For large values of  $\stheta$ at the upper end of the range reachable in phase~I (right panel), an energy upgrade may not be necessary, but a significantly longer baseline for the LENF in phase~II is preferable to determine the mass hierarchy: $1600 \, \mathrm{km} \lesssim L_2 \lesssim 4000 \, \mathrm{km}$. The energy upgrade (HENF) buys about 10\% in the fraction of $\deltacp$, where slightly shorter baselines than in the IDS-NF baseline are optimal: $2000 \, \mathrm{km} \lesssim L_2 \lesssim 4000 \, \mathrm{km}$.

For the intermediate case (middle panel), the CPV performance is significantly worse than for the high energy version. Therefore, an energy upgrade seems to be desirable, within about the same baseline range $2000 \, \mathrm{km} \lesssim L_2 \lesssim 4000 \, \mathrm{km}$. Note that the LENF in phase~II can measure the mass hierarchy for about 70\% to 90\% of all $\deltacp$ in the range $2000 \, \mathrm{km} \lesssim L_2 \lesssim 4000 \, \mathrm{km}$, which is below threshold in the figure.

Since $L_2 \simeq 4000 \, \mathrm{km}$ is sufficiently close to optimum, we choose that baseline together with the energy upgrade for the following discussion in phase~III, such as for CPV not found in phase~II.

\subsection{Phase~III: Third baseline or detector upgrade}

\begin{figure}[t]
\begin{center}
\includegraphics[width=0.45\textwidth]{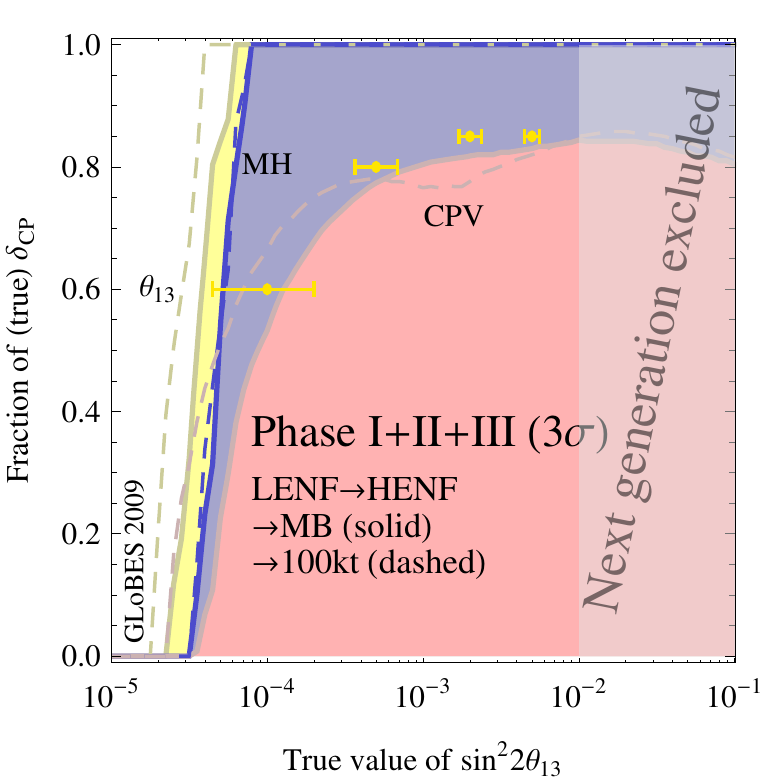}
\end{center}
\mycaption{\label{fig:resultP3} Results from phases  I+II+III. The different regions show the $\theta_{13}$, MH, and CPV discovery reaches ($3 \sigma$) as a function of $\stheta$ and fraction of $\deltacp$. The solid curves correspond to the baseline upgrade in phase~III (magic baseline), the dashed curves to a detector upgrade.  In addition, some error bars for $\theta_{13}$ are shown for several selected best-fit points, as they are obtained from phase~I+II. They are computed for the true $\deltacp=\pi/2$ (as an example) at the 90\% CL. Here a normal hierarchy is assumed.}
\end{figure}

Here we consider two possible further upgrades: Another baseline for the HENF, or a detector mass upgrade from 50~kt to 100~kt for the MIND detector at the same location at $L=4000 \, \mathrm{km}$.
If only one baseline is used, all muons are delivered to the respective storage ring. 
If a second baseline is added, the muons can be (almost) arbitrarily split between the two possibly simultaneously operating storage rings. We have tested a simultaneous operation of both baselines in phase~III with an equal splitting of the muons between the two storage rings, and we have tested an option with all muons in the new storage ring. Since the differences are moderately small, we choose the second option. Again, the data from phases~I and~II are added. For the second baseline, we consider the magic baseline $L=7500 \, \mathrm{km}$, because it is known to lead to optimal performances for all performance indicators for small $\theta_{13}$ in the combination with the shorter baseline~\cite{Kopp:2008ds}.

We show the results for phase~III in \figu{resultP3}. In this figure, the solid curves correspond to the baseline upgrade in phase~III (magic baseline), the dashed curves to a detector upgrade.  

\subsubsection{$\boldsymbol{\theta_{13}}$ not discovered in phase~I+II}
\label{sec:p3nd}

If $\theta_{13}$ is not discovered in phase~I+II, the best option to increase the $\theta_{13}$ is obviously  increasing the detector mass in phase~III (dashed curves in \figu{resultP3}). In this case, The $\theta_{13}$ and CPV reaches in the $\theta_{13}$ direction can be significantly improved in phase~III compared to phase I+II alone. That means that in the case of a $\theta_{13}$ discovery, also the chance to observe CP violation increases.

\subsubsection{$\boldsymbol{\theta_{13}}$ discovered in phase~I+II}

If $\theta_{13}$ is discovered in phase~I+II, we will know it with a certain precision before phase~III. This is illustrated by the error bars in \figu{resultP3}, which are shown for several selected best-fit points, as they could be obtained from phase~I+II. Based on these results, the best upgrade strategy can be chosen. Since the MH and $\theta_{13}$ can, in most cases, be easily measured, the only performance indicator to be optimized is the fraction of $\deltacp$ for which CP violation will be discovered. For instance, for $\stheta=0.002$ (best-fit), degeneracies are important, and the second baseline is the best strategy. This case also includes the possibility that $\stheta$ is discovered already in phase~I, but CPV has not been found in phase~I+II.

\begin{figure}[t]
\begin{center}
\includegraphics[width=\textwidth]{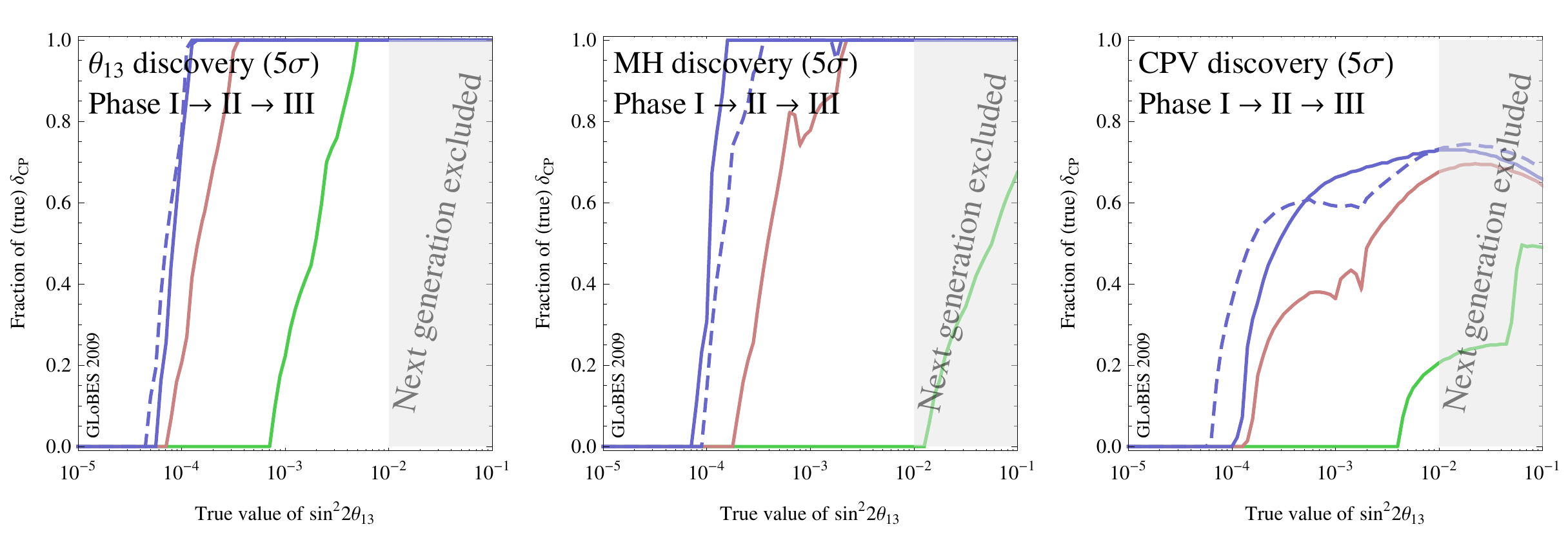}
\end{center}
\mycaption{\label{fig:phases} $\theta$, MH, and CP discovery reaches for a neutrino factory in three phases: Phase~I (light gray/green) is a low energy neutrino factory with a magnetized TASD, phase~II (medium gray/red) adds an energy upgrade with a MIND at the 4000~km baseline, and phase~III (dark gray/blue) includes another (magic) baseline (solid curves) or a detector upgrade at the 4000~km baseline (dashed curves). $5\sigma$ confidence level.}
\end{figure}

For $\stheta=0.0001$ (best-fit), however, the detector mass upgrade makes sense. For the other two cases shown in the figure, both options are equally good. Note that although the detector mass upgrade seems to be a good option
in many cases, the resolution of degeneracies happens through sufficient statistics. This means that if the target luminosity cannot be reached or a higher confidence level is chosen, the magic baseline is the more robust solution. We illustrate this in \figu{phases} below for the $5\sigma$ confidence level (where, for instance, the mass hierarchy discovery reach of the magic baseline option is superior).

Note that, in principle, one could also optimize the second baseline again for an optimal CPV sensitivity depending on the $\theta_{13}$ from phase~I and~II. However, here we focus on the magic baseline because it provides orthogonal, \ie, qualitatively different, physics, which may be also useful for non-standard measurements.

\vspace*{0.5cm}

We summarize the neutrino factory in stages for small $\stheta$ in \figu{phases} ($5\sigma$), where a second baseline is used in phase~III (solid curves) or the detector upgrade (dashed curves). At the $5\sigma$ confidence level, the strength of the magic baseline to resolve degeneracies becomes more pronounced, such as for the mass hierarchy discovery reach. In addition, there are many other applications of this baseline, see, \eg, \Refs~\cite{Winter:2004mt,deGouvea:2005mi,Gandhi:2006gu,Minakata:2006am}, for non-standard physics, see, \eg, \Refs~\cite{Ribeiro:2007ud,Kopp:2008ds}. However, the largest increase in the discovery reaches will come from the energy upgrade.

As far as the number of different storage rings is concerned, the LENF storage ring from phase~I cannot be recycled in phase~II since it points to a shorter baseline (for geometric reasons). Typically, it is also assumed
to be much smaller because of the smaller gamma factor. In phase~III a new storage is required if the magic baseline is chosen as option, whereas the detector upgrade does not require another storage ring. Therefore, either two or three different storage rings are needed in total.

\section{Summary and conclusions}

We have discussed the optimization of a low energy neutrino factory (LENF) and high energy neutrino factory (HENF), including the possibility of an energy upgrade from LENF to HENF. We have also pointed out that the magnetized TASD, which is proposed in the context of the LENF, might be used as an off-axis detector (OAD) in the HENF, which exactly the same beam spectrum if the off-axis angle is chosen accordingly. We have tested the impact of luminosity, baseline length, platinum channel, Daya Bay data (as representative for the next generation of experiments), additional baselines, and the matter density uncertainty on the absolute performance and optimization. Conceptually, we have distinguished the small $\stheta$ case ($\stheta \lesssim 0.01$) and the large $\stheta$ case ($\stheta \gtrsim 0.01$), which correspond to $\theta_{13}$ excluded or discovered by the next generation of experiments.

For large $\stheta$, we have demonstrated that a moderate luminosity LENF at a baseline of about 1100 to 1400~km may be the most plausible ``minimal effort'' neutrino factory option to measure CPV and MH. Because of the short baseline and relatively low energies, it is robust with respect to the knowledge of the matter density profile, and it allows for additions or upgrades, such as electron neutrino appearance (platinum channel), which significantly increase the sensitivity.  We have not found any significant impact on the performance if the data from a prior $\stheta$ measurement, such as from Daya Bay, are directly added. However, the  obtained rough value of $\stheta$ is helpful for the optimization (our optimization has been performed for certain true values of $\stheta$).

We have also discussed the HENF for large $\stheta$, which can basically perform the same measurements with a similar target sensitivity if the matter density can be controlled at a level below 2\%. In this case, the optimal baseline would rather be $2 000 \, \mathrm{km}$ to $4 000 \, \mathrm{km}$, depending on the value of $\stheta$ found and the matter density uncertainty. An OAD detector, which may also allow for the platinum channel because of the lower energies in the off-axis spectrum, can significantly enhance the sensitivity. The reason is that the T-inverted golden channel corresponds to the  platinum channel, which allows for a direct extraction of the intrinsic phase $\deltacp$ without extrinsic CP violation from the matter effect, compared to the CP-conjugated  golden channel (such as the antineutrino channel). While the optimal baseline for CPV is at about $1000 \, \mathrm{km}$ to $1500 \, \mathrm{km}$, baselines up to $4500 \, \mathrm{km}$ still allow for a CPV measurement for 80\% of all $\deltacp$.  Note that a HENF could be interesting for large $\stheta$ for different reasons, such as a case for new physics searches at the neutrino factory. For example, for non-standard matter effects, high muon energies are mandatory for sensitivities beyond the current limits~\cite{Kopp:2008ds}. In this case, also the $\tau$ production threshold is significantly exceeded, which allows for a detection of $\nu_\tau$.  For HENF+OAD, the baseline could be chosen in a wide window to search for new physics.

For small $\stheta$ ($\stheta$ not discovered by the next generation of experiments), we have plotted a possible staging scenario starting with a LENF (phase~I), followed by an energy upgrade (phase~II) and then either a detector mass upgrade or second (magic) baseline (phase~III). We have demonstrated that the LENF in phase~I could significantly enhance the $\stheta$ discovery reach beyond the exclusion limit expected from the next generation of experiments. The energy upgrade in phase~II has the largest impact on the sensitivities. Depending on the outcome of phase~I (if $\stheta$ is discovered), the baseline can be chosen accordingly, where larger values of $\stheta$ prefer slightly shorter baselines than smaller values of $\stheta$. We have also demonstrated that there is a synergy in resolving degeneracies between the LENF in phase~I and the HENF in phase~II. In phase~III, one can either upgrade the detector mass, or the baseline. The decision for either of the two options can be based on the outcome of phase I+II. The advantage of the detector mass upgrade is that it does not require a new (large inclination) storage ring, and a single-baseline neutrino factory can be effectively operated with a factor of two higher luminosity than a two-baseline neutrino factory, because the muons do not have to be split between the different storage rings. The benefit of the magic baseline option is qualitatively different physics, such as relevant for new physics searches in the presence of more than six oscillation parameters (see, \eg, \Ref~\cite{Ribeiro:2007ud}). We summarize the neutrino factory in stages for small $\stheta$ in \figu{phases} ($5\sigma$), where a second baseline is used in phase~III (solid dark gray/blue curves) or a detector upgrade (dashed dark gray/blue curves).

In summary, we have studied the optimization of a neutrino factory, where we have removed the constrained that the muon energy has to be fixed. We have found that for large $\stheta$, a neutrino factory including a TASD (either on- or off-axis) has very good performance. If high muon energies are needed,
such as to exceed the $\tau$ production threshold, a HENF should include an off-axis detector. For small $\stheta$, we have demonstrated that there can be a reasonable staging scenario including the LENF as first option, followed by an energy upgrade. We conclude that distinguishing LENF and HENF as clearly separate options may not be close to reality. A  realistic program may include components of both options, no matter if $\stheta$ is small or large.

\subsubsection*{Acknowledgments}

We would like to thank A. Bross, P. Huber, T. Li, and the members of the IDS-NF for useful discussions. Furthermore,
we would like to acknowledge support from the
Emmy Noether program of Deutsche Forschungsgemeinschaft,
contract WI 2639/2-1. This work was also supported by the 
European Union under the European Commission
Framework Programme~07 Design Study EUROnu, Project 212372.

\end{document}